\documentstyle[epsfig,12pt]{article}
\newcommand{\mycomm}[1]{\hfill\break
$\phantom{a}$\kern-3.5em{\tt===$>$ \bf #1}\hfill\break}
\newcommand{\mycommA}[1]{\hfill\break
$\phantom{a}$\kern-3.5em{\tt***$>$ \bf #1}\hfill\break}

\catcode`\@=11 
\let\rel@x=\relax
\newcount\timecount
\newcount\hours \newcount\minutes  \newcount\temp \newcount\pmhours
\hours = \time
\divide\hours by 60
\temp = \hours
\multiply\temp by 60
\minutes = \time
\advance\minutes by -\temp
\def\hour{\the\hours}
\def\minute{\ifnum\minutes<10 0\the\minutes
            \else\the\minutes\fi}
\def\clock{
\ifnum\hours=0 12:\minute\ AM
\else\ifnum\hours<12 \hour:\minute\ AM
       \else\ifnum\hours=12 12:\minute\ PM
            \else\ifnum\hours>12
                 \pmhours=\hours
                 \advance\pmhours by -12
                 \the\pmhours:\minute\ PM
                 \fi
            \fi
         \fi
\fi
}

 \def\monthname{\rel@x\ifcase\month 0/\or January\or February\or
   March\or April\or May\or June\or July\or August\or September\or
   October\or November\or December\else\number\month/\fi}

\def\bold#1{\setbox0=\hbox{$#1$}     \kern-.025em\copy0\kern-\wd0
     \kern.05em\copy0\kern-\wd0
     \kern-.025em\raise.0433em\box0 }

\def\lsim{\mathrel{\mathpalette\@versim<}}
\def\gsim{\mathrel{\mathpalette\@versim>}}
\def\@versim#1#2{\vcenter{\offinterlineskip
        \ialign{$\m@th#1\hfil##\hfil$\crcr#2\crcr\sim\crcr } }}
\catcode`\@=12

\begin{document}
\def\beq{\begin{equation}}
\def\eeq{\end{equation}}
\def\MSbar {\hbox{$\overline{\hbox{MS}}\,$}}
\def\eff{\hbox{\it\footnotesize eff}}
\def\APT{\hbox{\it\footnotesize APT}}
\def\mysim{\kern -.1667em\lower0.8ex\hbox{$\tilde{\phantom{a}}$}}
\vskip 20pt

\begin{titlepage}
\begin{flushright}
{\footnotesize
TAUP-2472-98\\}
\end{flushright}
\begin{centering}

{\large{\bf
Relations between Observables and the Infrared Fixed-Point in QCD
}}
\vskip 30pt

\vskip 30pt
{\bf Einan Gardi} \,\,\, and \,\,\, {\bf Marek Karliner} \\
\vspace{.05in}
School of Physics and Astronomy
\\ Raymond and Beverly Sackler Faculty of Exact Sciences
\\ Tel-Aviv University, 69978 Tel-Aviv, Israel
\\ e-mail: gardi@post.tau.ac.il, marek@proton.tau.ac.il

\vspace{0.9cm}
{\bf Abstract} \\
\vspace{0.35cm}
{\small
We investigate the possibility that $\alpha_s$ freezes as function of
$N_f$ within perturbation theory. We use two approaches  -- direct search
for a zero in the effective-charge (ECH) $\beta$ function, and the
Banks-Zaks (BZ) expansion.  We emphasize the fundamental difference
between quantities with space-like vs. those with time-like momentum.
We show that within the ECH approach several space-like quantities
exhibit similar behavior.  In general the 3-loop ECH $\beta$
functions can lead to freezing for $N_f\gsim 5$, but higher-order
calculations are essential for a conclusive answer.  The BZ expansion
behaves differently for different observables.  Assuming that the
existence of a fixed point requires convergence of the BZ expansion for
{\em any} observable, we can be pretty sure that there is no fixed point
for $N_f\lsim 12$.  The consequences of the Crewther relation
concerning perturbative freezing are analyzed.  We also emphasize that
time-like quantities have a consistent infrared limit only when the
corresponding space-like effective charge has one.  We show that
perturbative freezing can lead to an analyticity structure in the
complex momentum-squared plane that is consistent with causality. 
} 
\end{centering}

\end{titlepage}
\vfill\eject

\section{Introduction}

The running of the strong coupling constant $x(Q^2)=\alpha _s(Q^2)/\pi $ at
large momentum transfers $Q^2$ 
is well determined by the 1-loop perturbation theory result: 
\begin{equation}
x(Q^2)\sim \frac{\beta _0}{\ln \left( Q^2/\Lambda ^2\right) }
\label{1_loop_running}
\end{equation}
where 
\beq
\beta _0=\frac 14\left( 11-\frac 23N_f\right) 
\label{beta0}
\eeq
and $\Lambda $ is the QCD scale. Naively this leads to a divergent coupling
at $Q^2$ close to $\Lambda $ (a ``Landau pole''). However,
higher-order
terms, as well as non-perturbative effects are expected to alter the
infrared behavior and remove the divergence from any physical quantity.

It was suggested in the past, in various contexts 
\cite{MatSt}-\cite{Brodsky},
that infrared effects in QCD can be partially described using a coupling
constant that remains finite at low-energies. 
A finite coupling in the infrared limit can be achieved in many
ways without altering too much the ultraviolet behavior, which is 
well described by perturbation theory.
In the dispersive approach \cite{dispersive,Analytic}, for instance, 
the ``Landau-pole'' is removed by power corrections, and therefore the 
infrared coupling is
essentially non-perturbative. On the other hand, it is possible
\cite{BZ} that the coupling constant freezes
already within perturbation theory, due to a zero in the $\beta $ function 
induced by the 2-loop or higher-order corrections.
Such a perturbative infrared fixed point clearly occurs if the number
of light fermions is just below $N_f=16\frac12$ \cite{BZ}.

Phenomenological studies, such as \cite{MatSt} and references therein,
show that a
coupling-constant that freezes at low energies may be useful to
describe experimental data. But there is no general 
theoretical argument why QCD should lead to freezing.
On the contrary, the general belief, which is supported by lattice
simulations \cite{Lattice} and other approaches \cite{Appelquist}, is
that for small $N_f$ (and for 
pure Yang-Mills in particular) there is no infrared fixed point. One 
appears only for large enough number of fermions. 
Thus there is some critical $N_f^{crit}$ ($0<N_f^{crit}<16\frac12$)
such that a fixed point exists only for $N_f>N_f^{crit}$. 
The common lore 
is that below $N_f^{crit}$ the theory is in a confining phase, with 
spontaneously broken chiral symmetry \cite{Banks_Casher} and that  
the existence of an infrared fixed point is
related to the restoration of chiral symmetry \cite{Appelquist}.

The question of the infrared behavior of the coupling constant
immediately brings up the question of scheme dependence. The coupling
constant is in general not a physical quantity, and only the first two
coefficients of the $\beta$ function are scheme-independent. It is
clear, therefore, that the infrared stability of the coupling is
scheme-dependent, and so is its infrared limit value, when it exists.
Observable quantities can be used to define effective charges.
These are useful, because
contrary to arbitrary renormalization schemes, 
perturbative freezing of the coupling constant in physical schemes
can provide some indication of the existence of a fixed point in the full
theory. Clearly, the perturbative analysis can be trusted only if it
leads to freezing at small enough coupling constant values.

There are, in general, two techniques that have been used to study the
infrared limit of a generic observable within perturbative QCD. The
first is a direct search for a zero in the $\beta$
function, in a renormalization scheme that is `optimized' to
describe the particular observable \cite{MatSt,Ree_Kat,GLS_Kat,Higgs_fp}. 
The second is the Banks-Zaks (BZ) approach \cite{BZ,St,CaSt,Grunberg}.

The `optimized scheme' approach refers to a scheme in which one
hopes that higher
order corrections to the observable under consideration are 
small. In this case the fact that one is using a truncated series for
the observable and a truncated $\beta$
function is hopefully insignificant. In particular, two schemes have
been used to study freezing: the Principle of
Minimal Sensitivity (PMS) \cite{PMS}, and the method of
Effective Charges \cite{ECH}.  
A detailed analysis of the infrared behavior based on the
`optimized scheme'
methodology was conducted in ref. \cite{MatSt} for total hadronic
cross section in $e^+e^-$ annihilation ($R_{e^+e^-}$), in
ref. \cite{Ree_Kat} for $R_{e^+e^-}$ and for the $\tau$ lepton hadronic
decay ratio ($R_\tau$), in ref. \cite{GLS_Kat} for the 
Gross-Llewellyn Smith (GLS) sum rule for neutrino-proton scattering
and in ref. \cite{Higgs_fp} for the derivative of the hadronic decay
width of the Higgs.

The Banks-Zaks (BZ) approach \cite{BZ} 
originates in the observation that in a model where the
number of light quark flavors is just below $N_f^{*}=16\frac 12$ (which is
the critical value of $N_f$ at which $\beta _0$ changes sign), the
perturbative $\beta $ function (see eq. (\ref{beta}) below) 
is negative for very small values of the
coupling constant, but due to the 2-loop term it immediately 
crosses zero and becomes positive.
This perturbative infrared fixed point occurs at
$x_{FP}\simeq-1/c\equiv -\beta_0/\beta _1$, 
where $\beta_1$ is the 2-loop coefficient of 
the $\beta$ function\footnote{The $N_f$ dependence of $\beta_1$ is such that 
it is negative for $N_f^{**}<N_f<N_f^{*}$,
where $N_f^{**}\cong 8.05$}. $x_{FP} \longrightarrow 0$
as $N_f$ approaches $N_f^{*}$.
It was proposed \cite{BZ,Grunberg} 
to study higher-order effects on the fixed point
for models with a varying $N_f < N_f^{*}$, by expressing $x_{FP}$ 
as a power series in the ``distance'' from $N_f^{*}$, 
i.e. in $(N_f^{*}-N_f)$. It
was later suggested by Stevenson \cite{St} that this
perturbative freezing of the coupling constant might be relevant for the
real-world QCD with only two light flavors.

Recently the calculation of the 4-loop $\beta $ function \cite
{fourloops} has enabled Caveny and Stevenson \cite{CaSt} to check the
reliability of the BZ expansion for the location of the fixed point
in the physical renormalization schemes defined through the
effective charges of $R_{e^+e^-}$ and of the Bjorken polarized sum rule, 
by calculating the ${\cal O}\left((16\frac12-N_f)^3\right)$ order term  
in the corresponding BZ series. The authors of \cite{CaSt} have found
that the relevant BZ coefficients are small for both observables 
they considered. They suggested 
that the small coefficients indicate that the BZ expansion indeed 
holds for $N_f$ as
low as $2$, and therefore that the corresponding effective charges do
 indeed freeze  due to perturbative effects in the real-world QCD.
Caveny and Stevenson also investigated BZ expansion for 
the derivative of the Higgs hadronic decay width and for the anomalous
dimension that is defined from the derivative of the $\beta$ function at
the fixed point. 
They found that the coefficients of these series are rather large. 
This may indicate that the BZ expansion
breaks down at some larger $N_f$. 
This inconsistency was also one of the reasons we became 
interested in the subject.

Another method for obtaining a finite coupling in the infrared limit
from perturbation theory, which attracted much interest recently,
 is the so-called ``dispersive approach'' or
``analytic approach'' \cite{Analytic,dispersive}. 
The idea is to construct a time-like distribution by performing an
analytic continuation of the running coupling constant, and then use a 
dispersive integral to define an effective space-like coupling that is 
free from spurious singularities such as a ``Landau-pole''.

The main purpose of this paper is to examine all the evidence for the
existence of a perturbative fixed point in physical renormalization
schemes and its dependence on $N_f$. 
We use both the `optimized scheme' methodology and the BZ
approach and study several different observables in order to get a
global picture. We discuss relations between different physical
quantities, such as the Crewther relation \cite{Crewther}, 
and study their implication of perturbative freezing. 
We show and analyze the difference between quantities defined
with space-like momentum and those with time-like momentum.

This paper is composed of three main sections.
The first (Sec. 2) is devoted to the
`optimized scheme' methodology and the second (Sec. 3) -- 
to the BZ expansion. In third section (Sec. 4) we discuss
infrared behavior of time-like quantities. 
The main conclusion are given in Sec. 5.

The `optimized scheme' part starts with a brief introduction on 
scheme dependence and optimized schemes in QCD (Sec. 2.1), followed by
a computation and a discussion 
on the second renormalization group (RG) invariant ($\rho_2$)
for various time-like and space-like quantities (Sec. 2.2). 
In Sec. 2.3 we analyze the consequences
of the Crewther relation regarding perturbative freezing. In Sec. 2.4
we examine the numerical proximity in $\rho_2$ for several
space-like quantities. In Sec. 2.5 we analyze stability of the
fixed point in the $R_{e^+e^-}$ effective charge in the `optimized scheme'
approach. The main conclusions of Sec. 2 are given in Sec. 2.6.

The BZ section (Sec. 3) starts with a short introduction of the idea
and the essential formulae (Sec. 3.1). We then present the
BZ expansion in $\MSbar$ (Sec. 3.2), followed by
an analysis of the BZ expansion for time-like quantities (Sec. 3.3) and a
calculation of the BZ series for various quantities (Sec. 3.4).
The consequences of the Crewther relation are given (Sec. 3.5)
and finally we discuss the BZ expansion for the derivative of the
$\beta$ function at the fixed point (Sec. 3.6).
The conclusions from combining the analysis of Sec. 2 and Sec. 3 are
given in Sec. 3.7.   

In Sec. 4 we examine consistency of the perturbative approach and the
``analytic'' approach for describing
the infrared region of time-like observables, such as  $R_{e^+e^-}$. 
In Sec. 4.1 we shortly review the dispersive relations between the
vacuum polarization and $R_{e^+e^-}$. In Sec. 4.2 we examine the
analyticity structure of the D-function resulting from perturbation
theory and show that it can be consistent with the one expected from
causality only if the running coupling freezes. In Sec. 4.3 we 
examine the analytic perturbation theory approach.

\section{Fixed Points from `Optimized Schemes'}

\subsection{Scheme Dependence and `Optimized Schemes'}

We start by introducing the notation for the QCD $\beta$ function,
\begin{equation}
\beta (x)=Q^2\frac{dx}{dQ^2}=-\beta _0x^2-\beta _1x^3-\beta _2x^4-\cdots=-\beta
_0x^2\left( 1+cx+c_2x^2+\cdots \right)   
\label{beta}
\end{equation}
where $\beta_0$ is given in (\ref{beta0}), the two loop coefficients
is \cite{oneloop,twoloops}: 
\begin{equation}
c=\frac{\beta _1}{\beta _0}=\frac 1{4\beta _0}\left[ 102-\frac{38}
3N_f\right]   \label{c}
\end{equation}
and higher-order coefficients $c_2,c_3,\cdots$ 
depend on the renormalization scheme, and are given  
in $\MSbar$ by \cite{threeloops,fourloops}:
\beq
c_2=\frac 1{16\beta _0}\left[
  \frac{2857}2-\frac{5033}{18}N_f+\frac{325}{54} N_f^2\right] 
\label{c_2}
\eeq
\begin{eqnarray}
\label{c_3}
c_3 &=&\frac 1{64\beta _0}\left[  \left( \frac{149753}6+3564\zeta
_3\right) -\left( \frac{1078361}{162}-\frac{6508}{27}\zeta _3\right)
N_f  \right. \nonumber \\
&&+\left. \left( \frac{50065}{162}+\frac{6472}{81}\zeta _3\right)
  N_f^2+\frac{1093}{729}N_f^3\right]
\end{eqnarray}
where $\zeta_n$ is the Riemann zeta function ($\zeta _3\cong 1.202$, 
$\zeta_5\cong 1.03693$).

A generic physical quantity in QCD can be written in the form of 
an effective charge:
\beq
x^{\eff}=x\left(1+r_1 x +r_2x^2+r_3x^3+\cdots\right)
\label{x_eff}
\eeq
where $x=\alpha_s/\pi$ depends on the renormalization scheme and scale.
Specifying the expansion parameter $x$ amounts to choosing all the
coefficients of the $\beta$ function ($c_i$, $i\geq2$) and then
setting the renormalization scale.  

Consistency of the perturbative expansion (\ref{x_eff}), 
together with the RG equation (\ref{beta}), 
requires the invariance under RG of
the following quantities for a given QCD observable \cite{ECH}:
\beq
\rho = r_1-\beta_0\ln\left(\frac{Q^2}{\Lambda^2}\right)
\label{rho}
\eeq
at first order, and 
\begin{eqnarray}
\label{RS_invariants}
\rho_2& =& c_2+r_2-r_1^2-c_1 r_1 \nonumber\\
\rho_3& =& c_3+2 r_3+4 r_1^3+c_1 r_1^2-6 r_1 r_2-2 r_1 c_2 
\end{eqnarray}
at second and third orders, respectively. Similar quantities can be
defined at higher-orders \cite{ECH}.

At any finite order, a perturbative calculation has some residual
renormalization scheme dependence. In the infrared limit, the coupling
constant in different schemes can either diverge or freeze to a finite value.
Therefore, the very existence of an infrared 
fixed point in a perturbative finite order calculation for any QCD
quantity is scheme dependent. 
Of course, so is the value of the infrared coupling, when it is finite.

Although theoretically any scheme is legitimate, in practice the choice
of scheme is important even far from the infrared limit: there are
schemes in which the perturbative series (or the $\beta$ function series) 
diverges badly and there are schemes in which a finite order
calculation yields a good approximation to the physical value.
In the following we shortly review two special schemes 
-- the method of Effective Charges (ECH) \cite{ECH} and the Principle of
Minimal Sensitivity (PMS) \cite{PMS}. These schemes are `optimized', in
the sense that the coefficients of the $\beta$ function, as well as the
renormalization scale, are set in a way suited to
describe a specific QCD observable. As mentioned in the
introduction, it was conjectured in the past
\cite{MatSt,Ree_Kat,GLS_Kat,Higgs_fp} that perturbative calculations in these
schemes can be meaningful down to the infrared limit, although, of
course, the infrared values do not directly correspond to the 
measurable physical
quantities, which are governed by non-perturbative effects.
For $R_{e^+e^-}$ it was demonstrated \cite{MatSt} 
that if the experimental data are
``smeared'', they can be fitted by the perturbative result in the 
PMS\footnote{See Sec. 2.5 and Sec. 4 for a discussion on the applicability of
  the PMS/ECH methods to $R_{e^+e^-}$.} scheme down to low energies. 

The ECH method \cite{ECH} is based on using the actual observable effective
charge as a coupling constant: $x_{\small{\rm ECH}}\equiv x^{\eff}$. 
One way of achieving this is by choosing the renormalization scale and the
renormalization scheme, i.e. the coefficients of the $\beta$ function,
such that all the coefficients $r_i$
in eq. (\ref{x_eff}) are exactly zero\footnote{There are other choices
  possible at higher orders \cite{Ree_Kat}. 
For instance at the three-loop order one
  can choose $r_1+r_2x=0$. These different possible schemes coincide
  at the fixed point.}.
It is easy to see that in this scheme,
the coefficients of the $\beta$ function are simply the invariants
$\rho_i$ listed in (\ref{RS_invariants}): $c_2^{{\small {\rm ECH}}}=\rho_2$,  
$c_3^{{\small {\rm ECH}}}=\rho_3$, 
and so on. Thus, in order to find the value of the effective charge at the
fixed point in a finite order calculation in the ECH scheme, one just
looks for real positive solutions for the equation
$\beta^{\small{\rm{ECH}}}(x)=0$. For example, at the three-loop level one has:
\beq
1+cx^{\eff}+\rho_2\left(x^{\eff}\right)^2=0.
\label{ECH_FP}
\eeq
One further requirement is, of course, that $x^{\eff}$ at the fixed
point will be small enough, so that the perturbative
expansion will be trustworthy down to the infrared limit.
It is clear that if $c$ is {\em negative} and in particular if the $c$ term
dominates over the $\rho_2$ term, there will be an infrared fixed
point at approximately  $x^{\eff}\sim -1/c=-\beta_0/\beta_1$. 
This is indeed the case in QCD with
$N_f$ just below $N_f=16\frac12$ which is the critical value at which
$\beta_0=0$. This is the starting point for the Banks-Zaks approach,
to which we shall come back in Sec. 3.
However, the physically relevant values for $c$ are always positive
($c$ is positive for $N_f<8.05$). Then a real positive solution for
eq. (\ref{ECH_FP}) can only originate from a negative $\rho_2$.

The PMS method \cite{PMS} is based on using a renormalization scale and scheme
which is the least sensitive to a local change in the RG
parameters. For instance, at the three-loop order, $x^{\eff}$ depends
on two free parameters, which can be chosen to be the coupling $x$ and
$c_2$. Thus the renormalization scheme dependence of $x^{\eff}$ can be
studied from the shape of the two dimensional surface of 
$x^{\eff}=x^{\eff}(x,c_2)$. The PMS scheme corresponds to a saddle
point on this surface at which
\beq
\frac{\partial x^{\eff}}{\partial x}=0
\eeq
and
\beq 
\frac{\partial x^{\eff}}{\partial c_2}=0.
\eeq
The resulting equation, based on the condition 
$\beta^{{\small {\rm PMS}}}(x)=0$, is \cite{MatSt}:
\beq
\frac74+c\, x_{{\small{\rm PMS}}}+3\left(\rho_2-\frac{c^2}{4}\right)
x_{{\small{\rm PMS}}}^2=0
\label{PMS_FP}
\eeq
and after solving the equation for $x_{{\small{\rm PMS}}}$ one
substitutes it in (\ref{x_eff}) to get the value of $x^{\eff}$ at the
fixed point.
Here, like in the ECH case, eq.  (\ref{PMS_FP}) has
a real and positive solution for a positive $c$ and a negative
$\rho_2$. However, unlike the ECH case, here there is still a window for a
small positive $\rho_2$ ($0< \rho_2<c^2/4$), in which eq. (\ref{PMS_FP}) has a
positive solution. In practice $c$ is relatively small,
and the conditions for the existence of a fixed point at
the three-loop order are very similar in the ECH and PMS schemes.
For both the ECH and PMS methods, the perturbative fixed point from
the three-loop order analysis occurs at small coupling, provided
$\rho_2$ is {\em large  and negative}.
It was also found in \cite{MatSt,Ree_Kat,GLS_Kat} that the value of
$x^{\eff}$ at the freezing point is somewhat lower in the PMS
method than it is in the ECH case.  

\subsection{The Second RG-invariant for Time-like and Space-like Quantities}

We saw that the existence of a perturbative fixed point in the ECH/PMS
approach at the three-loop order depends crucially on the sign of
$\rho_2$. The fixed point found this way can only be considered
reliable if it occurs at small enough value of the coupling -- and this
in turn depends on the magnitude $\rho_2$.
Therefore, the calculation of $\rho_2$ is the first step in
analyzing the perturbative infrared behavior of a QCD effective charge
at the three loop level.

In fig. 1 we present $\rho_2$ as a function of $N_f$ for several
QCD observables:    
\begin{description}
\item{1)}
The Bjorken sum rule for the deep inelastic scattering of polarized
electrons on polarized nucleons, defined by 
\beq
\int_0^1
dx\left[g_1^{ep}(x,Q^2)-g_1^{en}(x,Q^2)\right]\equiv\frac16\vert
g_A\vert \left[1-\frac{\alpha_{Bj}}{\pi}\right]
\eeq
where
\beq 
\frac{\alpha_{Bj}}{\pi}\equiv x_{Bj}=x(1+k_1x+k_2x^2+\cdots)
\label{k_series}
\eeq
with the perturbative result at NNLO from ref. \cite{PBjSR_NNLO}.
\item{2)}
The Bjorken sum rule for deep-inelastic neutrino nucleon
scattering, 
\beq
\int_0^1
dx\left[F_1^{\overline{\nu} p}(x,Q^2)-F_1^{\nu n}(x,Q^2)\right]
\equiv 1-\frac{C_f}{2}\left(\frac{\alpha_{F_1}}{\pi}\right).
\eeq
where
\beq 
\frac{\alpha_{F_1}}{\pi}\equiv x_{F_1}=x(1+f_1x+f_2x^2+\cdots)
\label{f_series}
\eeq
with the perturbative result at NNLO from ref. \cite{F1_NNLO}.
\item{3)}
The Gross-Llewellyn Smith sum rule (GLS) for neutrino proton scattering,  
\beq
\int_0^1dx
\left[F^{\overline{\nu}p}_3(x,Q^2)+F^{\nu p}_3(x,Q^2)\right]\equiv
6\left[1-\frac{\alpha_{GLS}}{\pi}\right]
\eeq
where
\beq 
\frac{\alpha_{GLS}}{\pi}\equiv x_{GLS}=x(1+l_1x+l_2x^2+\cdots)
\label{l_series}
\eeq
with the relation to the coefficients of the 
polarized Bjorken sum rule given by \cite{PBjSR_NNLO}:
\begin{eqnarray}
\label{light_by_light}
l_1&=&k_1\\ \nonumber
l_2&=&k_2-\frac{d^{abc}d^{abc}}{C_f N_c}\left(-\frac{11}{144}
+\frac16\zeta_3\right)N_f
\end{eqnarray}
where the difference between the 3-loop coefficients of the two
observables in (\ref{light_by_light}) is due to the light-by-light 
type diagrams.
\item{4)}
The vacuum polarization D-function (not a directly measurable
quantity) defined as the logarithmic derivative of the vector current
correlation function $\Pi(Q^2)$, with a space-like momentum  $Q^2=-q^2>0$,
\beq
4\pi^2 i\int d^4x e^{iq\cdot x}\left\langle 0 
\vert T\left\{j^\mu(x),j^\nu(0)\right\}\vert 0 \right\rangle = 
(q^\mu q^\nu-q^2g^{\mu\nu})\Pi(Q^2)
\label{Pi_def}
\eeq
\beq
D(Q^2)=Q^2\frac{d\Pi(Q^2)}{dQ^2}
=3\left(\sum_fQ_f^2\right)\left[1+\frac{\alpha_D}{\pi}\right]
\label{D_def}
\eeq
where
\beq 
\frac{\alpha_{D}}{\pi}\equiv x_{D}=x(1+d_1x+d_2x^2+\cdots)
\label{d_series}
\eeq
with the perturbative result at NNLO from ref. \cite{Ree_NNLO}.
In (\ref{D_def}) we ignored the contribution from the light-by-light type
diagrams which is
proportional to $(\sum_f Q_f)^2$. The corresponding diagrams contribute
to $x_D$ starting at three-loops: $\Delta x_D=d_2^{lbl}x^3+O(x^4)$ . 
We are interested in studying a purely QCD
phenomenon, and therefore it is inconvenient to include these terms which
involve also the electromagnetic interaction.
However, it is still interesting to see whether this
neglected contribution influences our conclusions concerning the infrared
behavior.
We will study this issue indirectly by comparing the results for the Bjorken
polarized sum rule to those for the GLS 
sum rule, since ({\sl cf.} (\ref{light_by_light}))
\beq
d_2^{lbl}=(l_2-k_2)\frac{(\sum_f Q_f)^2}{N_f}.
\eeq 

\item{5)}
The total hadronic cross section in $e^+e^-$ annihilation (again
neglecting the light-by-light terms),
defined by
\beq
R(s)\equiv 3\left(\sum_fQ_f^2\right)\left[1+\frac{\alpha_R}{\pi}\right]
\label{R_def}
\eeq
where
\beq 
\frac{\alpha_{R}}{\pi}\equiv x_{R}=x(1+r_1x+r_2x^2+\cdots)
\label{r_series}
\eeq
The perturbative coefficients of $R_{e^+e^-}$ 
can be related to those of the vacuum polarization D-function, 
by using the dispersion relation (see Sec. 4). 
The relations are \cite{Bjorken}:
\begin{eqnarray}
\label{r_d_relations}
r_{1}&=&d_{1}\nonumber\\
r_{2}&=&d_{2}-\frac{\pi^2\beta_0^2}{3}\nonumber\\
r_{3}&=&d_{3}-\pi^2\beta_0^2 \left(d_1+\frac56c\right).
\end{eqnarray}
For our purpose, it is convenient to  
write the relations between the corresponding 
RG invariants $\rho_i$ defined in (\ref{RS_invariants}):
\begin{eqnarray}
\label{rho_r_D_relations}
\rho_2^R&=&\rho_2^D-\frac13\pi^2\beta_0^2 \\ \nonumber
\rho_3^R&=&\rho_3^D-\frac53\pi^2\beta_0^2 c
\end{eqnarray}

\item{6)}
The $\tau$ lepton hadronic decay ratio $R_{\tau}$, defined by \cite{tau}, 
\beq
R_{\tau}\equiv
\frac{\Gamma(\tau^-\rightarrow \nu_\tau\,\mbox{hadrons} \, (\gamma))}
{\Gamma(\tau^-\rightarrow\nu_{\tau}e^-\overline{\nu}_e(\gamma))} 
=3\left(1+\frac{\alpha_{\tau}}{\pi}\right).
\eeq
where
\beq 
\frac{\alpha_{\tau}}{\pi}\equiv x_{\tau}=x(1+\tau_1x+\tau_2x^2+\cdots)
\label{tau_series}
\eeq
and where $(\gamma)$ indicates possible presence of photons in the final state.
 The perturbative
coefficients of $r_{\tau}$ are also related to those of the vacuum 
polarization D-function, as follows \cite{tau}:
\begin{eqnarray}
\label{tau_d_relations}
\tau_1&=&d_1-\beta_0 I_1 \\ 
\nonumber
\tau_2&=&d_2-(2d_1+c)\beta_0I_1+\beta_0^2I_2 \\ 
\nonumber
\tau_3&=&d_3-(3d_2+2d_1c+c_2)\beta_0I_1+
\left( 3d_1+\frac{5}{2}c\right)\beta_0^2I_2-\beta_0^3I_3 
\end{eqnarray}
where $I_1=-19/12$, $I_2=265/72-\pi^2/3$ and $I_3=-3355/288+19\pi^2/12$.    

\item{7)}
The static potential, defined by \cite{Peter}
\beq
V(q^2)=-C_f\frac{4\pi\alpha_V(q^2)}{q^2}
\eeq
where $q^2$ is the three-momentum squared, corresponding 
to the spatial separation $r$ between the quark and the anti-quark,
and where 
\beq
\frac{\alpha_V}{\pi} \equiv x_V=x(1+v_1x+v_2x^2+\cdots).
\eeq
The static potential was recently calculated \cite{Peter} up to order 
${\cal O}(\alpha_s^3)$.

\item{8) }
The derivative of $\Gamma_H$, the Higgs hadronic decay width, defined by
\cite{Higgs}
\beq
\frac{\alpha_H}{\pi} \equiv x_H
=-\frac12\frac{d\ln\left(\Gamma_H/M_H\right)}{d\ln{M_H}^2}
=x(1+h_1x+h_2x^2+\cdots)
\eeq
where $M_H$ is the Higgs mass, which is assumed to be much larger
than the quark masses.
$x_H$ was recently calculated \cite{Higgs} up to order 
${\cal O}(\alpha_s^4)$. This means that $\rho_3$, the
4-loop coefficient of the ECH $\beta$ function is now available. 
\end{description}

The observations from fig. 1 are:
\begin{description}
\item{a)}
There is a clear distinction between the time-like quantities and the
space-like quantities.
\item{b)}
There is a surprising numerical proximity 
between $\rho_2$ for several different space-like 
quantities: this includes the vacuum polarization D-function, the GLS
sum rule and the polarized and non-polarized Bjorken sum rules, but
{\em not} the static potential. The proximity 
is particularly evident for $N_f\leq5$. 
This issue and its relation to the
ideas of ref. \cite{CSR} are further discussed in Section 2.4.
\item{c)} For the space-like
quantities (except the static potential) 
$\rho_2$ becomes positive for $N_f\lsim4$, and thus 
according to the ECH/PMS approach at this order there is no
fixed point for these values of $N_f$.
\item{d)} The static potential behaves differently from the other
  space-like quantities. $\rho_2^V$ becomes positive already for
  $N_f\lsim 9$.
\item{e)} At low $N_f$,
 $\rho_2^R$, $\rho_2^{\tau}$ and $\rho_2^D$ are 
numerically very different from one another. 
For larger $N_f$ they become closer. At $N_f>8$, the $\rho_2$
values for the three effective charges are large and negative
and very close to each other\footnote{This is due to their 
relation through an analytic continuation as discussed in Sec.
4.}, indicating that they freeze to similar values (see fig. 2).
\item{f)} For the time-like quantities $\rho_2$ stays
negative down to $N_f=0$ or $1$, indicating a possible infrared fixed
point according to the ECH/PMS approach. 
\item{g)} $\rho_2^H$ ($\rho_2$ for the derivative of the Higgs decay
  width) behaves differently from the others: it is not a
  monotonically decreasing function of $N_f$, and it is negative and
  large for any $N_f$. 
\item{h)} The second coefficient of the $\beta$ function in $\MSbar$ ($c_2$)
  is not close to those of the effective charge schemes ($\rho_2$).
\end{description}

Let us briefly discuss the question of freezing of $x_H$, the
effective charge defined from the derivative of the Higgs decay width,
as we believe it can 
teach us a general lesson. Naively, the fact that $\rho_2^H$ is negative and
large means that there is an infrared fixed point at a
  rather small $x_H$ value. For instance, for $N_f=3$, $\rho_2^H\simeq
  - 57.6$, and thus the ECH $\beta$ function (see eq. (\ref{ECH_FP})) has
  a zero at $x_H^{FP}\simeq 0.15$.  Now, since the value of $x_H^{FP}$
  is small, one could further conclude that the perturbative analysis is
  reliable. But is this really the case?
This issue was discussed in detail in
  ref. \cite{Higgs_fp}, where is was conjectured that this fixed point
  is spurious, and that the perturbative series breaks down
  at NLO in this case.
In general, the fact that $\rho_2 x^2$ becomes equal to the leading terms
$1+cx$ does not immediately imply breakdown of the perturbative
series\footnote{If that was the case, then `perturbative
  freezing' would have been a meaningless term.}. 
It is still possible that $\rho_3 x^3$ (and maybe a few higher
order terms) will be smaller than  $\rho_2 x^2$, while the
asymptotic nature of the series will take over at some higher order 
(for a recent review see \cite{Fischer}).  
 Since the next order term for $x_H$ is now available \cite{Higgs} 
this question can be clarified explicitly: it turns out that indeed the
4-loop ECH $\beta$ function 
$-\beta_0 x^2\left(1 + cx + \rho_2^H x^2 + \rho_3^H x^3 \right)$ 
has a real and positive zero only for $N_f\gsim 14$.
For $N_f \lsim 13$ the term $\rho_3^H x^3$ turns out to be larger 
than $\rho_2^H x^2$ {\em before}  $\rho_2 x^2$ becomes larger than the
leading terms $1+cx$. This confirms that the conjecture in
ref. \cite{Higgs_fp} that the fixed point in $x_H$ (for a small $N_f$) is a
spurious one and that the series breaks down at NLO.
It is clear that all-order resummation 
is essential in this case. The general lesson is that one should 
exercise extreme caution when looking for a fixed point in a finite
order calculation.
The NNLO analysis of the existence of a fixed point in the ECH/PMS
schemes is based on the assumption that
 $\left\vert \rho_3 x_{FP}\right\vert<\left\vert\rho_2\right\vert$,
which may turn out to be wrong.

Now we proceed to consider other quantities: 
in fig. 2 we present the ECH value of the effective
charges at freezing obtained through solving eq. (\ref{ECH_FP}), 
for the D-function and for the related
time-like quantities: $R_{e^+e^-}$ and $R_{\tau}$. 
Of course, this calculation can only be 
meaningful if $\rho_2<0$. From (\ref{ECH_FP}) it is clear that $x_{FP}^{\eff}$
diverges as $\rho_2 \longrightarrow 0^-$, corresponding to $N_f \simeq
4$ for $x_D$ and to $N_f\simeq 1$ for $x_\tau$.

The $x_D$ result in fig. 2 actually represents very well also the
results for the other space-like quantities (except the one for the static
potential $x_V$). The reason is transparent from figs. 1 and 3 -- the
latter showing the differences between the values of $\rho_2$ for
various quantities and $\rho_2^D$: the
differences between $\rho_2^D$, $\rho_2^{Bj}$, $\rho_2^{F_1}$
and $\rho_2^{GLS}$ are rather small in the region 
$N_f\leq8$, where the three loop contribution is important for
freezing. For higher value of $N_f$, freezing is induced by the
two loop $\beta$ function, which is invariant, and the
differences in $\rho_2$ almost do not alter the value of the effective
charge at freezing.  

In fig. 2 we show the PMS result for the D-function only. 
The PMS results for the effective charge at freezing are somewhat
lower than the ECH result (in accordance with \cite{MatSt,Ree_Kat}) 
but the general picture is the same. 

Purely perturbative effective charges, at any
order in perturbation theory, can in principle 
diverge in the infrared, independent
of whether or not the full theory has an infrared fixed point. 
A priori, it is also possible for different effective charges 
to have a totally different perturbative infrared behavior for a
given $N_f$. 
In particular, it is possible that
there will be a ``Landau-pole'' in one effective charge $x_a$, while 
another effective charge $x_b$ will exhibit perturbative freezing. 
Still one could write a power expansion of $x_b$ in term of $x_a$
(and vice-versa) \cite{CSR,Gen_Crewther}.
One would then expect such an expansion to be divergent.  
However, in practice the numerical proximity between the $\rho_2$ coefficients 
for several different space-like
effective charges suggests that the expansion of $x_a$ in terms of $x_b$
is at least close to being convergent (See Section 2.4). 
It seems that the different
space-like effective charges considered above (excluding $x_V$) 
are so closely related that perturbative 
freezing could only occur for all of them together or -- for none.

\subsection{Freezing and the Crewther Relation} 

The only example where one can explicitly verify our conjecture
about simultaneous perturbative freezing of 
different quantities is the Crewther relation \cite{Crewther}.
This is an all-order relation between
the vacuum polarization and the polarized Bjorken sum rule. In terms of
effective charges it can be written as
\beq
x_{Bj}-x_D+\frac34 C_f x_{Bj}x_D=-\frac13 \beta(x) S(x)
\label{crewther}
\eeq
where the $\beta$ function is defined in (\ref{beta}), 
$S(x)$ is a power series in the coupling constant
\beq
S(x)=S_1+S_2x+S_3x^2+\cdots
\eeq
and $S_i$ depend on $N_f$ and $N_c$.
Writing the effective charges as power series (eqs. (\ref{k_series})
and (\ref{d_series})) one obtains a relation between the 
coefficients $k_i$ of $x_{Bj}$ and $d_i$ of $x_D$, as follows:
\begin{eqnarray}
\label{coef_crewther}
k_1&=&d_1-\frac34C_f+\frac13\beta_0S_1  \nonumber \\ 
k_2&=&d_2-\frac34C_f(k_1+d_1)+\frac13(\beta_0S_2+\beta_1S_1)
 \nonumber\\
k_3&=&d_3-\frac34C_f(d_1k_1+d_2+k_2)
+\frac13(\beta_0S_3+\beta_1S_2+\beta_2S_1)
\end{eqnarray}
From the knowledge of the 3-loops coefficient of the Bjorken sum rule
\cite{PBjSR_NNLO} 
and the vacuum polarization \cite{Ree_NNLO} we can obtain $S_1$ and $S_2$:
\beq
S_1 = - \frac{21}{2} + 12 \zeta_3
\label{S_1}
\eeq
and
\beq
S_2 = \frac{221}{3} \zeta_3 C_a -\frac{629}{8} C_a -\frac{38}{3} N_f
\zeta_3+
\frac{397}{24}C_f+34C_f\zeta_3-60C_f\zeta_5+\frac{163}{12}N_f
\label{S_2}
\eeq
where $S_2$ is scheme dependent and is given here in $\MSbar$.
 
The term on the r.h.s. of eq. (\ref{crewther}) is scheme-invariant,
but $\beta(x)$ and $S(x)$ are separately scheme dependent. If $x_D$
has a perturbative fixed point $x_D^{FP}$, 
then it is convenient to write the r.h.s. of
(\ref{crewther}) in terms of $x_D$. $\beta(x_D^{FP})=0$ and so
 the r.h.s. vanishes
at $x_D=x_D^{FP}$. Therefore $x_{Bj}$ also freezes perturbatively,
leading to the original conformal Crewther relation:
\beq
x_{Bj}^{FP}=\frac{x_D^{FP}}{1+\frac34 C_f x_D^{FP}}.
\label{crewther_conformal}
\eeq
The argument works, of course, 
in both directions, i.e. if the Bjorken effective 
charge freezes, then the D-function will also freeze to the value:
\beq
x_{D}^{FP}=\frac{x_{Bj}^{FP}}{1-\frac34 C_f x_{Bj}^{FP}}.
\label{crewther_conformal_2}
\eeq
Note that (\ref{crewther}) allows a situation in which {\em both}
$x_D$ and $x_{Bj}$ diverge in the infrared limit.

We shall use the Crewther relations in the following section, where we
study the numerical proximity of $\rho_2$ between various space-like 
quantities, and also later, in conjunction with the BZ approach.

\subsection{Numerical proximity of $\rho_2$ for different space-like
  quantities}

In this section we study the numerical proximity between
the $\rho_2$ invariants for the various space-like quantities (fig. 1)
mentioned in Section 2.2.
The values of $\rho_2^{Bj}$, $\rho_2^D$, $\rho_2^{F_1}$ and $\rho_2^{GLS}$
are given in Table 1, for  $N_c=3$ and $N_f\leq7$. 
In fig. 3 we show the difference between
the values of $\rho_2$ for the various space-like quantities and
$\rho_2^D$. The vertical scale is enlarged here by a factor
of $10$ with respect to that in fig.~1.  
\vskip20pt

\begin{table}
\centerline{
\begin{tabular}{||c||c|c|c|c||}
\hline
$N_f$&$\rho_2^{Bj}$ &$\rho_2^D$&$\rho_2^{F_1}$&$\rho_2^{GLS}$\\
\hline
\hline
$0$ & 17.812  &     17.924      &   15.956 &  17.924 \\
\hline
$1$ & 13.557  &      13.747     &   12.344 &  13.334 \\
\hline
$2$ &  9.371  &       9.605     &   8.702  &  8.779\\
\hline
$3$ &  5.237   &       5.475    &   5.007  &  4.236 \\
\hline
$4$ &  1.134   &       1.330    &   1.223  & -0.322 \\
\hline
$5$ & -2.969   &    -2.869      &    -2.691& -4.935 \\
\hline
$6$ & -7.114   &     -7.177     &   -6.798 & -9.657 \\
\hline
$7$ &-11.361   &   -11.669      &  -11.186 & -14.562\\
\hline
\end{tabular}
}
\caption{$\rho_2$ for various space-like quantities for $N_f\leq7$
\label{tab}}
\end{table}

The numerical proximity  between $\rho_2$ for 
the GLS sum rule and $\rho_2$ for the Bjorken
polarized sum rule is obvious, as they differ only by 
the light-by-light diagrams at 3-loops level, giving rise to a 
small\footnote{It is interesting that this difference becomes important
  in the framework of the BZ expansion, as we discuss in Sec. 3.4.} 
contribution, proportional to $N_f$. 
This difference of course vanishes
exactly for $N_f=0$. 
Therefore, we focus on the three remaining quantities:
the vacuum polarization D-function, and the polarized and non-polarized 
Bjorken sum rule.

One's initial guess is to suspect that the similarity between $\rho_2$ for 
the Bjorken polarized sum rule and $\rho_2$ for the vacuum
polarization D-function is due the Crewther relation (see Sec. 2.3). 
We show below, however, that this numerical agreement is a second 
``miracle'', on top of the Crewther relation.

The numerical agreement between $\rho_2^{F_1}$ and
the rest of the space-like quantities at low $N_f$
is not as good:
at $N_f=0$ the difference is about $10\%$,
vs. $0.6\%$  relative difference between $\rho_2^{Bj}$ and $\rho_2^D$. 
Nevertheless, for larger $N_f$
$\rho_2^{F_1}$ is quite close to the others.

We now investigate further the numerical proximity
of $\rho_2^{Bj}$ and $\rho_2^D$, for two reasons:
first, the numerical agreement in this case is remarkable
for $0\leq N_f \leq 7$ ({\sl cf.} Table 1)\footnote{The agreement
  is not as good for larger $N_f$ values, for instance, for $N_f=14$,
  $\rho_2^{Bj}= -80.2$ while $\rho_2^D= -68.8$, about $14\%$
  difference. This is to be compared with a difference of
$1\%$ to $3\%$ for $N_f\leq3$.}, 
second, it is interesting to see to what extent this numerical proximity
is related to the Crewther relation.

From (\ref{RS_invariants}) and (\ref{coef_crewther}) we get:
\beq
\rho_2^{Bj}-\rho_2^{D}=\frac14C_f \beta_0 S_1+\frac13\beta_0S_2
-\frac23 d_1\beta_0S_1-\frac19\beta_0^2S_1^2+\frac34c\,C_f
\label{D_Bj_diff}
\eeq
where $d_1$ is defined in (\ref{d_series}) 
and $S_1$ and $S_2$ are given in (\ref{S_1})
and (\ref{S_2}). Note that both $d_1$ and $S_2$ depend on the
renormalization scheme and scale, but in such a way that  
$\rho_2^{Bj}-\rho_2^{D}$ is scheme and scale invariant.
Substituting the three-loop expressions into (\ref{D_Bj_diff}) 
we obtain a complicated
function of $N_c$ and $N_f$, with no clue that
$\rho_2^{Bj}-\rho_2^{D}$ is small.
To make things simple, we first consider the case $N_f=0$:
\begin{eqnarray}
\label{D_Bj_diff_num}
\left. \rho_2^{Bj}-\rho_2^{D}\right|_{N_f=0}&=&N_c^2\left(-\frac{1043}{4752}
-\frac{869}{108}\zeta_3-\frac{55}{6}\zeta_5+\frac{121}{9}\zeta_3^2\right)\\
\nonumber
&&-\frac{23053}{19008}-\frac{253}{36}\zeta_3+\frac{55}{6}\zeta_5
\end{eqnarray}
Numerically, we have:
\beq
\left. \rho_2^{Bj}-\rho_2^{D}\right|_{N_f=0} \simeq 0.02966 N_c^2-0.15542
 \eeq
The ``miracle'' is in the numerics! In (\ref{D_Bj_diff_num}) one finds
all the irrational numbers that enter the three loop calculation
$\zeta_3$, $\zeta_3^2$ and  $\zeta_5$. Each of the terms separately
is of order 1 to 10, but they combine to give a tiny sum. 

It is important to note that there is nothing special in
the $N_f=0$ case considered above. 
In order to get a more general view of the relative magnitude of difference 
$\rho_2^{Bj}-\rho_2^D$, 
we consider the normalized difference
${\cal R}$
defined by:
\beq
{\cal R}= \frac{\large{\vert}\rho_2^{Bj}-\rho_2^D\large{\vert}}
{\large{\vert}\rho_2^{Bj}\large{\vert}+\large{\vert}\rho_2^D\large{\vert}}
\label{Rcal}
\eeq
$\cal R$ is plotted in fig. 4 as a function of $N_c$ for various
values of $N_f$, and in fig. 5, as a function of $N_f$, for various
values of $N_c$. In
general, {\em $\cal R$ is of order $1\%$! } 
The only occasions where the relative difference ${\cal R}$ is not
small (the peaks raising above ${\cal R}=0.1$ in fig. 4 and 5) is
when both $\rho_2^{Bj}$ and $\rho_2^D$
are close to zero. Then they may even have opposite signs, leading to
${\cal R}=1$.

We conclude that the numerical proximity between $\rho_2^D$ and
$\rho_2^{Bj}$ is not a direct consequence of the factorization implied
by the Crewther relation (r.h.s. in eq. (\ref{crewther})), but of the
particular numerical coefficients. While the numerical proximity of 
$\rho_2^{GLS}$ and $\rho_2^{Bj}$ is well understood, we do not know of
any reason why $\rho_2^{F_1}$ is close to $\rho_2^{Bj}$.  
It is tempting to think that there is a deeper reason for the
numerical proximity, and that higher order ECH coefficients 
($\rho_i$ for $i>2$) for different quantities are also close.
In this respect it will also be interesting to know the fundamental
reason why $\rho_2^V$ is so different than $\rho_2$ for the other
space-like quantities.  

Next we discuss the relation between the assumption that 
$\rho_i$ for different observables are
close to one another, and the work by Brodsky and Lu \cite{CSR}
on commensurate-scale relations between observables.
In \cite{CSR} (see also \cite{Gen_Crewther}) it was 
suggested to express one effective charge ($x_b$) in terms
of another ($x_a$), and then choose the scale of $x_a$ according to the
BLM criterion \cite{BLM}. In \cite{CSR} it was found that the
coefficients in the expansion $x_b=x_b(x_a)$ 
are both much simpler than the ones in
some arbitrary scheme (like $\MSbar$), and are {\em numerically
  small}. Let us see what are the conditions for the coefficients in
such an expansion to be small. We start with the expressions for 
two generic effective charges in some scheme:
\begin{eqnarray}
\label{xa_xb}
x_a=x+r_1^a x^2+r_2^a x^3+\cdots \\ \nonumber
x_b=x+r_1^b x^2+r_2^b x^3+\cdots
\end{eqnarray}
and express $x_b$ in terms of $x_a$:
\beq
x_b=x_a+m_1{x_a}^2+m_2{x_a}^3+\cdots
\label{ba}
\eeq 
where $m_i$ depend on $r_i^a$ and $r_i^b$, for instance: 
\hbox{$m_1=r_1^b-r_1^a$}, 
\hbox{$m_2=r_2^b-r_2^a-2r_1^a(r_1^a-r_1^b)$}. Note that $m_i$ are, by
definition, invariant with respect to the choice of the intermediate
renormalization scheme. The next step is to use the
definitions of the RG invariants
(\ref{RS_invariants}), and express the coefficients of (\ref{ba}) 
in terms of $\rho_i^a$ and $\rho_i^b$. $m_1$ is just the difference
between the values of the first invariant $\rho$ (\ref{rho}) 
for the two effective charges:
\beq
m_1=r_1^b-r_1^a=\rho^b-\rho^a\equiv \Delta,
\eeq
and
higher-order coefficients can be expressed entirely in terms of
$\Delta$ and of the higher-order invariants $\rho_i^a$ and $\rho_i^b$,
and {\em depend mainly on $\Delta$ and on the differences of 
the coefficients $\rho_i$}. 
We define
$\delta_i=\rho_i^b-\rho_i^a$, and then,
\begin{eqnarray}
m_2&=&\delta_2+\Delta^2+c\Delta\\ \nonumber
m_3&=&\frac12\delta_3+\Delta(\rho_2^b+2\delta_2)+\frac52c\Delta^2+\Delta^3 
\end{eqnarray}
$\Delta$ can be tuned by changing the scale of $x_a$,
while the scale of $x_b$ is kept fixed. In particular, there are
choices of scale for which $\Delta$ is small, such as the
leading-order BLM scale \cite{BLM,CSR}, which
eliminates all $\beta_0$ terms from $\Delta$, or, simply the 
  choice $\Delta=0$ \footnote{An
  advantage of the first over the latter is that in the first the scale
  does not depend on the number of light flavors, and thus observables
  cross the quark thresholds together. This issue is discussed in
  detail in ref. \cite{CSR}.}. 
A small $\Delta$ is, however,
not enough to guarantee small higher-order $m_i$ coefficients. The
latter will be small only if
$\delta_i$ are also small.

One practical conclusion from this discussion is that relating 
$x_D$, $x_{Bj}$, $x_{GLS}$ and $x_{F_1}$ to one another as suggested
in \cite{CSR} will probably lead to more accurate results than some
generic scheme. However, relating any of the above to $x_V$ is disfavored.

This concludes our discussion on the numerical proximity of the
$\rho_2$ coefficients for space-like effective charges. Next,
we briefly discuss the possibility of applying the 
`optimized-scheme' approach to study the infrared limit of time-like
effective charges.  

\subsection{The reliability of the ``fixed point'' in $R_{e^+e-}$ from ECH}

In this section we study further the `optimized scheme' approach
applied directly to the $R_{e^+e^-}$ effective charge, along the lines 
of ref. \cite{MatSt,Ree_Kat}. We consider specifically the case
$N_f=3$ at the three- and four-loops order, 
and try to estimate the reliability of the ECH analysis in the infrared. 
A deeper study of time-like quantities in the context of freezing 
is postponed to Sec. 4.
 
\setcounter{footnote}{0}
The `optimized scheme' approach was found to be unreliable 
when applied directly to
time-like quantities in another context, by Kataev and Starshenko
\cite{KatStr}. They found that the ECH and PMS
methods for estimating the next term in a series\footnote{These
methods are based on the assumption that the next, 
uncalculated coefficient in the
ECH/PMS $\beta$ function, is close to zero} 
when applied directly to two- and three-loop series for 
time-like quantities such as $R_{e^+e^-}$, do not predict the
correct structure of the terms that result from the analytic
continuation (the situation is similar for Pad\'e approximants). 
The solution of \cite{KatStr} is to predict the
coefficients of the space-like D-function and use the exact relations 
(\ref{r_d_relations})
to obtain the coefficients of the time-like quantity. 
One will naturally expect that if
PMS and ECH fail in predicting the next terms when applied directly to
the time-like quantities, they should not trusted for studying the
infrared limit of these quantities. 
Nevertheless, it may still be instructive to see what one obtains in
this approach.

Considering eq. (\ref{rho_r_D_relations}) one finds that 
the $\pi^2$ terms, that make the coefficients of the ECH $\beta$
function of $x_R$ ($\rho_i^R$) different from those of $x_D$
($\rho_i^D$),  are 
numerically significant and negative. This is
the reason why the analysis in Sec. 2.2 indicates an infrared fixed point 
for the $x_R$ and not for $x_D$ for $N_f\leq 4$. 

Taking as an example QCD with $N_f=3$, we examine the ECH $\beta$ function
for the D-function and $R_{e^+e^-}$: 
\beq
\beta=-\beta_0x^2(1+cx+\rho_2x^2+\rho_3x^3+\cdots).
\eeq
The two loop coefficient is invariant: $c=1.778$. The three loop
coefficients are $\rho_2^D \simeq 5.23$ and 
$\rho_2^R \simeq -11.42$, and the four loop coefficients are 
\beq
\rho_3^D \simeq -33.39+2d_3
\label{rho_3_D}
\eeq 
and 
\beq
\rho_3^R\simeq -181.98 + 2 d_3
\label{rho_3_R}
\eeq
where $d_3$, the $\MSbar$ four loop coefficient of the D-function, is
unknown. 

Next, we consider different approaches to predict $d_3$, and thus
$\rho_3^D$ and $\rho_3^R$.
In the ECH/PMS approaches for predicting the next term in a
perturbative series \cite{ECH,PMS,KatStr} one {\em assumes}
$\rho_3^D\simeq 0$ in
order to obtain a prediction for $d_3$. Therefore, one cannot use 
an ECH/PMS prediction to calculate $\rho_3^D$.
Pad\'e Approximants (PA)
\cite{PA_QCD} applied in $\MSbar$ 
predict $d_3=24.75$ (using the $[1/1]$ PA) and 
$d_3=16.49$ (using the $[0/2]$ PA). 
We note that these predictions are close to one another, and are also
consistent with the ECH/PMS assumption $\rho_3^D\simeq 0$, which leads
to $d_3=16.7$\footnote{The authors of \cite{KatStr} obtained a sightly
  larger value: $d_3=27.5$. The difference is due to the additional
  assumption taken in \cite{KatStr} that $c_3=0$ -- the 4-loop
  coefficient of the $\beta$ function was not known then.}.
One could alternatively try to use either PMS/ECH directly for the 
$R_{e^+e^-}$ (i.e. assume $\rho_2^R=0$, instead of
assuming $\rho_2^D=0$) or apply the PA's method directly for the $R_{e^+e^-}$ 
  series. However, as we already mentioned, 
the resulting predictions do not agree 
between the different methods and do not contain the
  correct $\pi^2$ terms \cite{KatStr} and are therefore not 
reliable.

We conclude that $\rho_3^D\simeq 0$ and $\rho_3^R\simeq -150$. Thus,  
while the question of whether the D-function effective charge
freezes for $N_f=3$ remains open, it is quite clear that the naive
`optimized scheme' approach predicts that $R_{e^+e^-}$ does:
$\rho_3^R$ is large and negative.
Nevertheless, if the value of $x_R$ at the fixed point 
is calculated (as the zero of the four-loop ECH $\beta$ function) using a
reasonable guess for $d_3$, one obtains a $x_R^{FP} \simeq
0.19\pm0.03$
\footnote{The only real solution of the equation 
$\beta^{{\rm ECH}}=0$ at 4-loops (with $N_f=3$) is pretty stable:
 it changes in the range \hbox{$0.16 \leq x_R^{FP}\leq 0.23$} 
for \hbox{$-20 \leq d_3\leq 60$}.} 
which is compared in fig. 2 with the three-loop ECH result
$x_R^{FP} \simeq 0.38$ (or the three-loop PMS results $x_R^{FP} \simeq
0.30$). 
We therefore conclude that the ECH (or PMS) methods fail in predicting 
the infrared limit of the perturbative result in this case. The
reasons for this will become clear in Sec. 4. 

A remark is in order concerning the Higgs decay width effective
charge $x_H$, which is also a time-like observable, and therefore contains
$\pi^2$ terms that result from the analytical continuation. 
In principle, the problem discussed above should appear in this case
as well. However, contrary
to the case of $x_R$ or $x_\tau$, the numerical significance of these
terms in the $x_H$ series is rather small 
(compared to other contributions at the same order) and therefore our
previous conclusions concerning $x_H$ hold. 

\subsection{Conclusions}

We analyzed here the freezing of the QCD effective charge for
various quantities by looking for a zero in the corresponding 
 ECH (or PMS) $\beta$ function.

We found a clear distinction between time-like and space-like quantities.
For several space-like quantities we found that the numerical values of 
the second RG invariant are quite close, especially
for a small $N_f$. This suggests that the different quantities are
closely related. It is tempting to conjecture that this will be
reflected in numerical proximity of $\rho_3$ and higher-order
coefficients of the corresponding ECH $\beta$ functions.
We also expect that when perturbative freezing occurs,  
it will occur together for the various quantities. This
is clearly true for the D-function and the Bjorken polarized sum rule,
due to the Crewther relation. For other quantities, such as the
Bjorken non-polarized sum rule, this is just a conjecture.

For the space-like quantities (except for the static potential) it
seems possible that there is a fixed point for $N_f\gsim 5 $, 
while there is no indication of freezing below $N_f\lsim5$. 
Absence of a perturbative fixed point at the three-loop level does not
necessarily mean that it does not exist. It simply means that one
should consider higher-order correction to answer this question. 
On the other hand, presence of a fixed point at the three-loop level,
does not guarantee that it will persist at higher orders.

In Sec. 3 we look for more clues for or against the relevance of
perturbative freezing, by studying the Banks-Zaks 
expansion of these quantities.

For time-like quantities, a naive application of the PMS/ECH approach
indicates a perturbative fixed point even when the corresponding
space-like quantity does not freeze.
On the other hand we identified an instability of the predicted
effective charge value at freezing. 
This result is a first indication of the inconsistency of the
approach, as discussed further in Sec. 4.

\section{The Banks-Zaks Expansion Approach}

\subsection{The BZ fixed point}

We start this section by summarizing the basics of the BZ
expansion \cite{BZ,CaSt,Grunberg} for the location of the fixed point. 

As is mentioned in the introduction, the BZ expansion is an
expansion in \hbox{$(N_f^*-N_f)$}, i.e. the ``distance'' from the 
critical value $N_f^{*}=16\frac 12$
down to lower values of $N_f$. It is convenient to use the expansion
parameter \cite{Grunberg,CaSt}:
\beq
a_0=\frac 8{321}\left( 16\frac 12-N_f\right) =\frac{16}{107}\beta _{0.}
\label{a0}
\eeq
Since $a_0$ is linear in $N_f$, it is straightforward to rewrite the $\beta $
function coefficients $\beta _i=\beta _0c_i$ as polynomials in $a_0$, 
obtaining ({\sl cf.} (\ref{beta0}), (\ref{c})-(\ref{c_3})):
\beq
c=-\frac{1}{a_0}+c_{1,0}=-\frac 1{a_0}+\frac{19}4
\label{c_a0}
\eeq
and 
\beq
c_2=c_{2,-1}\frac 1{a_0}+c_{2,0}+c_{2,1}a_0
\label{c_2_a0}
\eeq
\beq
c_3=c_{3,-1}\frac 1{a_0}+c_{3,0}+c_{3,1}a_0+c_{3,2}a_0^2
\label{c_3_a0}
\eeq
where explicit expressions for $c_{i,j}$ in $\MSbar$ 
are given in ref. \cite{CaSt}. The next
step is to solve the equation $\beta (x)=0$, yielding an expression for
$x_{FP}$. With the 2-loop $\beta $ function, one obtains:
\beq
x_{FP}=-\frac 1c=\frac{a_0}{1-\left( \frac{19}4\right) a_0}
\label{1_over_c}
\eeq
From (\ref{1_over_c}) it is clear that $x_{FP}$ 
is asymptotically proportional to $a_0$,
and therefore we look for higher-order solutions $x_{FP}$, 
in the form of a power expansion in $a_0$:
\beq
x_{FP}=a_0(1+u_1a_0+u_2a_0^2+\cdots).
\label{x_FP}
\eeq
Assuming that a fixed point exists, one then substitutes (\ref{x_FP})
in the equation $\beta (x)=0$, finding that $u_1$ is fully 
determined by the 3-loop coefficient of
the $\beta $ function, $u_2$ -- by the 4-loop
coefficient, etc. The formulae for the first three $u_i$-s are:
\begin{eqnarray}
\label{u_1_3}
u_1&=&c_{1,0}+c_{2,-1} \nonumber\\
u_2&=&c_{1,0}^2+2c_{2,-1}^2+3c_{1,0}c_{2,-1}+c_{2,0}+c_{3,-1} \nonumber \\  
u_3& =& 3 c_{2,0} c_{1,0}+4 c_{2,0} c_{2,-1}+4 c_{1,0}
c_{3,-1}+c_{1,0}^3+6 c_{1,0}^2 c_{2,-1} \nonumber \\ 
&&+10 c_{1,0} c_{2,-1}^2 
 +c_{4,-1}+5 c_{2,-1} c_{3,-1}+5 c_{2,-1}^3+c_{2,1}+c_{3,0}
\end{eqnarray}
$u_1,u_2,\cdots$ depend on the renormalization scheme. 
One may be interested in studying the fixed point
in some other scheme, related to the original one by
\beq
y=x(1+r_1x+r_2x^2+\cdots).
\label{y_x}
\eeq
In particular, one may be interested in the fixed point in some physical
scheme, in which case $y$ is an effective charge corresponding to
some measurable quantity. In order to obtain the appropriate expansion for
the location of the fixed point $y_{FP}$ in the new scheme, one first writes
the coefficients $r_i$ as polynomials in $a_0$: 
\beq
r_i=r_{i,0}+r_{i,1}a_0+\cdots+r_{i,i}a_0^i.
\label{r_i_a0}
\eeq
Next, one substitutes (\ref{r_i_a0}) and 
(\ref{x_FP}) in eq. (\ref{y_x}), obtaining
\beq
y_{FP}=a_0(1+w_1a_0+w_2a_0^2+\cdots)
\label{w_i}
\eeq
where
\begin{eqnarray}
\label{w_1_3}
w_1&=&u_1+r_{1,0} \nonumber \\
w_2&=&u_2+2r_{1,0}u_1+r_{2,0}+r_{1,1}\nonumber \\
w_3&=&u_3+2r_{1,0}u_{2}+2r_{1,1}u_{1}+
3r_{2,0}u_{1}+r_{1,0}u_{1}^2+r_{2,1}+r_{3,0} 
\end{eqnarray}
It is important to note that the coefficients $w_i$ for a given 
effective charge $y$ are free from any renormalization scheme
ambiguities \cite{Grunberg}, as scheme dependence  
cancels out between the $u_i$ and $r_i$ terms. This is expected on
general grounds, since both $y_{FP}$ and $a_0$ in eq. (\ref{w_i}) are
physical quantities.
This invariance can also be understood considering another possibility of
obtaining the BZ coefficients: as a first step one calculates the $\rho_i$
renormalization scheme invariants (\ref{RS_invariants}) of the
required effective charge. One then writes the coefficients $\rho_i$
as series in $a_0$, similarly to (\ref{c_2_a0}) and
(\ref{c_3_a0}), with the only difference that here also
$\rho_{i,i}$ terms appear, for instance:
\beq
\rho_2=\rho_{2,-1}\frac{1}{a_0}+\rho_{2,0}+\rho_{2,1}a_0+\rho_{2,2}a_0^2
\eeq
Finally, one calculates the BZ expansion for the value of $x_{FP}^{\eff}$,
directly from the ECH $\beta$ function
$\beta(x^{\eff})=-\beta_0\left(x^{\eff}\right)^2
\left[1+cx^{\eff}+\rho_2\left(x^{\eff}\right)^2
+\rho_3\left(x^{\eff}\right)^3+\cdots\right]$,
similarly to the way eq. (\ref{u_1_3}) was obtained for the $\MSbar$
$\beta$ function. The resulting BZ series is exactly equal to
the one obtained in (\ref{w_1_3}) using the BZ expansion in $\MSbar$
as an intermediate step.

\subsection{The BZ Expansion in $\MSbar$}

Before using the BZ approach to study perturbative freezing for
effective charges we consider the expansion in $\MSbar$, where the
coefficients of the $\beta$ function are given by eqs. (\ref{beta0}) and
(\ref{c}) through (\ref{c_3}). We are interested the range 
$0\leq N_f < 16\frac12$, in which there is asymptotic freedom. 
We note that $c$ is negative (leading to a fixed point at the two loop
order) only for $N_f>8.05$; $c_2$ is negative only for $N_f>5.84$, and
$c_3$ is never negative. This makes it clear that the $\MSbar$
coupling is not expected to freeze within the 4-loop calculation for 
$N_f\leq5$. 

In this situation we expect the BZ expansion to
break down at $a_0\simeq 0.26$ (corresponding to $N_f=6$).  
The coefficients of the BZ expansion 
are obtained directly from eq. (\ref{u_1_3}), 
\begin{eqnarray}
\lefteqn{x_{FP}^{\tiny \MSbar} =   a_0 +   \frac {11675}{10272} \,
  {a_0}^2 + \,\left(  \frac {145645559}{
17585664}  +   \frac {5335}{428} \,\zeta (3)\right){a_0}^3} \\
 & &  + \,\left( - {  \frac {
92177206455497}{1083839643648}}  - {  \frac {
587191201}{13189248}} \,\zeta (3) + c_{4,-1} \right){  a_0}^{4}
\nonumber
\end{eqnarray}
and numerically,
\beq
x_{FP}^{\tiny \MSbar} = {  a_0} + 1.1366\,{  a_0}^{2} + 23.2656\,
{  a_0}^{3} + \left( - 138.5630 +  c_{4,-1} \right) {a_0}^4.
\label{BZ_MSbar}
\eeq
We see that the series is indeed ill behaved.
For example, at $a_0=0.26$,  
the first three terms are roughly: $0.26$, $0.0768$, $0.409$.
In fact, the series seems to break down well above $N_f \sim 6$, but
using only three terms it is difficult to estimate exactly where.

We conclude that at least for $N_f\leq 6$
 the $\MSbar$ coupling
does not freeze, and that the ``Landau pole'' behavior of the 
1-loop coupling in the
infrared, persists in $\MSbar$ when higher-order corrections are
taken into account. But is this conclusion true to all orders?
We think that the answer is positive:
since the BZ expansion in $\MSbar$ seems to break-down already at
the order ${\cal O}(a_0^3)$, it is hard to imagine how higher-order
corrections could alter the situation. 
Still, it would be interesting to know how the BZ series behaves at
higher orders.

One approach \cite{Gracey} is to use the information that can be obtained on
the  $\MSbar$ $\beta$ function at higher-orders is from the
large-$N_f$ limit.  
From the ${\cal O}(1/N_f)$ terms calculated in \cite{Gracey} 
one can obtain all the 
${\cal O}(1/a_0)$ terms, denoted by $c_{i,i-1}$ ({\sl cf.}
(\ref{c_a0}) - (\ref{c_3_a0})). In ref. \cite{Gracey} it
was found that they are small, and that their effect on
the value of the fixed point (and thus also on the question of its
existence) is small.    

Another possibility is to use PA's \cite{PA_QCD} to estimate higher-order
coefficients, and use them to calculate higher-order terms in the BZ
expansion. Such an effort is now under way \cite{BZ_PA}.

\subsection{BZ Expansion for Time-like Quantities}
 
In Sec. 2 we found that the `optimized scheme' approach does not give
reliable results for the effective charge at freezing when applied 
directly to time-like quantities. 
As we show in Sec. 4, a consistent description of freezing for
$R_{e^+e^-}$ or $R_\tau$ exists only when the space-like D-function
freezes, and then the infrared limits are expected to be the same
for all three quantities: 
\beq
D(0)=R_{e^+e^-}(0)=R_{\tau}(0).
\label{ir_equality}
\eeq
This is a direct consequence of the expected analyticity structure of
the D-function, which indeed holds when perturbative freezing occurs.

The `optimized scheme' approach, in general, does not obey this
requirement, since the terms that are related to the analytical
continuation from space-like to time-like momentum, change the ECH (or 
PMS) $\beta$ function (see eq. (\ref{rho_r_D_relations})). 
Fig. 2 shows that the resulting difference between the values of the
effective charges at freezing are insignificant above
$N_f=8$. However, for $N_f\lsim 8$ these differences become
larger. This can be interpreted as a sign that the three-loop 
results are not
conclusively indicating freezing for these cases, which is true, but it 
is also related to the unjustified use of the ECH/PMS methods directly 
for time-like quantities.

The BZ approach, on the other hand,  yields
the {\em same} expansion for the three effective charges, the D-function, 
$R_{e^+e^-}$ and $R_\tau$, just as one expects from (\ref{ir_equality}). 
It is straightforward to check this:
from the relations between the coefficients 
of $R_{e^+e^-}$ (\ref{r_d_relations}) (or $R_\tau$
(\ref{tau_d_relations})) 
and those of the D-function, 
it is easy to obtain the relation between the 
corresponding coefficients of the different
powers of $a_0$ for the two quantities, namely $r_{i,j}$ and $d_{i,j}$
(or $\tau_{i,j}$ and $d_{i,j}$). 
The substitution of $r_{i,j}$ (or $\tau_{i,j}$) in terms of $d_{i,j}$
 into the formulae for the BZ $w_i$ 
coefficients (eq. (\ref{w_1_3})), gives the same
$w_i$ coefficients as for the D-function. All the terms that result
from the analytic continuation just cancel out between the different  
contributions, both for $R_{e^+e^-}$ and for $R_{\tau}$.

\subsection{BZ Expansion for various quantities}

In this section evaluate the coefficients in the BZ series for the 
fixed point in physical effective charges schemes for 
the various quantities discussed in Sec. 2.2. 

The calculation is straightforward using eq. (\ref{w_1_3}) and the
three loop coefficients from 
refs. \cite{PBjSR_NNLO,F1_NNLO,Ree_NNLO,Higgs,Peter}.
Thus we go directly to the results, followed by a discussion.
In the next section we study the implications of the Crewther relation between
$x_D$ and $x_{Bj}$ on the BZ expansion. Note that
the ${\cal O}(a_0^4)$ coefficients cannot be explicitly  
calculated due to the lack of both the 5-loop coefficients of the
$\beta$ function and the 4-loop coefficients of the series for
the different observables.
 
\begin{description}
\item{1) }
For the polarized Bjorken sum rule:   
\begin{eqnarray}
\label{BZ_BjSR}
\lefteqn{x_{FP}^{Bj}={  a_0} + {  \frac {753}{3424}} \,{  a_0}
^{2} + \left({  \frac {5930095}{17585664}}  -
{  \frac {275}{214}} \,\zeta _3\right)\,{  a_0}^{3}} \\
 & &  + \left( -   \frac {18602593666427}{
361279881216}  - {  \frac {9470237}{137388}} \,\zeta
 _3  - {  \frac {535}{8}} \,\zeta _5 +  c_{4,-1}  +  
k_{3,0} \right)\,{  a_0}^{4}
\nonumber
\end{eqnarray}
where $k_{3,0}$ is the 4-loop Bjorken polarized sum rule coefficient at 
$N_f=16\frac12$, as in the general definition in eq. (\ref{r_i_a0}).  
The numerical results are:
\beq
x_{FP}^{Bj}={  a_0} + 0.2199\,{  a_0}^{2} - 1.2075\,
{  a_0}^{3} + \left( - 203.6939 +  c_{4,-1} 
 +  k_{3,0} \right)\,{  a_0}^{4}
\label{PBjSR_num}
\eeq
\item{2) }
For the non-polarized Bjorken sum rule:
\begin{eqnarray}
\lefteqn{x_{FP}^{F_1}={  a_0} - {  \frac {4589}{10272}} \,{  a_0
}^{2} + \left({  \frac {425842061}{52756992}}  +
{  \frac {30809}{963}} \,\zeta _3 - { 
\frac {805}{18}} \,\zeta _5\right)\,{  a_0}^{3}} \\
 & &  + \left( - {  \frac {38886699582523}{
361279881216}}  - {  \frac {327376259}{2198208}} \,
\zeta _3 + {  \frac {2966545}{61632}} \,
\zeta _5  +    c_{4,-1}  + f_{3,0} \right)\,{  a_0}^{4}
\nonumber
\end{eqnarray}
or
\beq
x_{FP}^{F_1}={  a_0} - 0.4467\,{  a_0}^{2} + 0.1551\,{ 
a_0}^{3} + \left( - 236.7461 +  c_{4,-1} + f_{3,0} \right)\,{  a_0}^{4}
\eeq

\item{3) }
For the GLS sum rule:
\begin{eqnarray}
\lefteqn{x_{FP}^{GLS}=a_0+\frac{753}{3424}a_0^2+
\left(\frac{239442925}{52756992}-\frac{3355}{321}\zeta_3\right)a_0^3}\\
&&+\left(-\frac{5706068695529}{120426627072}-\frac{535}{8}\zeta_5
-\frac{171229447}{2198208}\zeta_3+c_{4,-1}+l_{3,0}\right)a_0^4
\nonumber
\end{eqnarray}
or
\beq
x_{FP}^{GLS}={  a_0} + 0.2199\,{  a_0}^{2} \,-8.02495{ 
a_0}^{3} + \left( - 210.3609 +  c_{4,-1} + l_{3,0} \right)\,{  a_0}^{4}
\eeq

\item{4) }
For the vacuum polarization D-function (and therefore, according to
Sec. 3.3, also for $R_{e^+e^-}$ and $R_\tau$):
\begin{eqnarray}
\label{Ree}
\lefteqn{x_{FP}^{D}={  a_0} + {  \frac {4177}{3424}} \,{  a_0}
^{2} + \left({  \frac {31250575}{17585664}}  -
{  \frac {275}{214}} \,\zeta _3\right)\,{  a_0}^{3}} \\
 & &  + \left({  \frac {81595375713359}{
1083839643648}}  - {  \frac {3893665183}{13189248}}
\,\zeta _3 + {  \frac {2675}{24}} \,
\zeta _5   + c_{4,-1} +d_{3,0} \right)\,{  a_0}^{4}
\nonumber
\end{eqnarray}
or
\beq
x_{FP}^{D}=
{  a_0} + 1.2199\,{  a_0}^{2} + 0.2323\,{ 
a_0}^{3}
 + ( - 164.0075 +  c_{4,-1} +d_{3,0})\,{  a_0}^{4}
\label{Ree_num}
\eeq

\item{5) } For the static potential:
\begin{eqnarray}
\lefteqn{x_{FP}^{V}=a_0 -   \frac {8869}{10272} \,a_0^2 + 
\,\left(  \frac {70824311}{17585664}
  +   \frac {27}{8} \,\pi ^{2} -   
\frac {9}{64} \,\pi ^{4} -   \frac {275}{214} \,
\zeta_3\right)\,a_0^3} \\ \nonumber
 & & + \,\left( -   \frac {130549250005577}{1083839643648}  
-   \frac {322462723}{3297312} \,\zeta_3
+ c_{4,-1} +v_{3,0}
-  \frac {105075}{219136} \,\pi ^{4}  \right.\\ \nonumber& & \left.
+  \frac {315225}{27392} \,\pi ^{2} 
\right)\,a_0^{4}
\end{eqnarray}
or
\beq
x_{FP}^{V}=a_0 - 0.8634\,a_0^{2} + 22.0945\,a_0^{3}
 +\,(- 171.1353+ c_{4,-1} +v_{3,0})\, a_0^{4}
\eeq

\item{6) }
For the derivative of the Higgs hadronic decay width: 
\begin{eqnarray}
\lefteqn{x_{FP}^H=a_0 +   \frac {31363}{10272} \,
a_0^2 + \,\left(  \frac {486174653}{52756992}  
-   \frac {275}{214} \,\zeta_3\right)\,a_0^3 } \\ \nonumber
 & &  + \,\left(  \frac {4675}{48} \,\zeta_5  - 
  \frac {294627948398435}{3251518930944}  - 
  \frac {982216871}{9891936} \,\zeta_3+c_{4,-1}\right)\,a_0^{4}
\end{eqnarray}

\beq
x_{FP}^{H}=a_0 + 3.0533\,a_0^{2} + 7.6707\,a_0^{3}
 +\,( - 108.9778 + c_{4,-1})\, a_0^{4}
\eeq
\end{description}

The conclusions are:
\begin{description}
\item{a) }
The BZ coefficients for $x_{Bj}^{FP}$, $x_{F_1}^{FP}$ and $x_{D}^{FP}$, 
 are of order $1$ up to ${\cal O}(a_0^3)$. Estimation of 
the  ``radius of convergence'' from the first few terms is difficult, but
convergence of these series is not ruled out for any positive $N_f$.
\item{b) }
The BZ coefficients for  $x_{GLS}^{FP}$, $x_{V}^{FP}$ and 
$x_{H}^{FP}$ are relatively large.
For  $x_{GLS}^{FP}$ and $x_{V}^{FP}$ it is very difficult to estimate 
the ``radius of convergence'' from the first few terms since $w_1$ coefficient
is especially small, while $w_2$ is rather large. 
For $x_{H}^{FP}$ it seems that the series converges for $a_0\lsim
0.36$, corresponding to $N_f\gsim 2$.
\item{c) }
Clearly, $w_2$ for $x_{GLS}^{FP}$ is large due to the three loop 
light-by-light term that makes the only
difference ({\sl cf.} (\ref{light_by_light})) between the  GLS sum rule and the
Bjorken polarized sum rule, for which $w_2$ is small. It may be
surprising at first sight that light-by-light contribution which
usually just a
small correction to the three loop invariant $\rho_2$ can make such a
difference for the BZ expansion. The reason is basically the fact that
the BZ expansion works around $N_f=16\frac12$, where the
light-by-light term is not small, being proportional to $N_f$.
In the vacuum polarization D-function, we neglected a similar
light-by-light type term. As explained in Section 2.2, the correction
expected in the D-function is smaller by a factor of $(\sum_f
Q_f)^2/N_f$, compared to the GLS sum rule case. This term is
therefore not expected to break the BZ expansion for the D-function.  
\item{d) } There is a noteworthy numerical cancelation  between
  terms containing different irrational and transcendental numbers,
 that is responsible for the small $w_2$ values.
This is particularly evident for the 
non-polarized Bjorken sum rule, where the rational term is roughly 
$8.07$, the term
proportional to $\zeta_3$ is $38.45$, and the term proportional
to $\zeta_5$ is $-46.37$, bringing the sum to about $0.155$!
Note that the case of the non-polarized Bjorken sum rule is 
special both in the fact
  that the ${\cal O}(a_0^3)$ coefficient in the BZ expansion already contains
  a $\zeta_5$ term, and in large cancelation between irrational
  numbers that occurs there.
\item{e) } In $x_H^{FP}$, which is a time-like quantity, 
all the $\pi^2$ terms that are related to the analytical continuation
  from space-like to time-like momentum that appear in the
  perturbative coefficients $h_i$ cancel out in the formula for the BZ
  series. This is in accordance with Sec. 3.3\footnote{Note that $\pi^2$ and
  $\pi^4$ terms do appear in the BZ expansion for $x_V^{FP}$ but these
  terms are not related to any analytical continuation.}.
 \end{description}

\subsection{The BZ Expansion and the Crewther Relation}

In this section we study the consequences of the 
Crewther relation \cite{Crewther} (see section 2.3) for the BZ
expansions for the Bjorken sum rule and the vacuum polarization
D-function.

The first observation is that the Crewther relation can be used to
evaluate the difference between the yet unknown  ${\cal O}(a_0^4)$ coefficients
in the BZ expansion of the two observables.
This observation is based on the fact that 
from the third relation in (\ref{coef_crewther}) one can obtain
$k_{3,0}-d_{3,0}$, the {\em exact} the difference between the 4-loop
coefficients of the Bjorken sum rule and the vacuum polarization
function at $N_f=16\frac12$: by 
substituting $\beta_0=0$ the only unknown, namely $S_3$, is
eliminated.  The result is:
\beq
k_{3,0}-d_{3,0}=
\frac{3439187}{27648}-\frac{257719}{1152}\zeta_3+\frac{535}{3}\zeta_5
\simeq 40.393
\label{k0_d0_diff}
\eeq
Knowing $k_{3,0}-d_{3,0}$ we look again at eqs.  (\ref{PBjSR_num})
and (\ref{Ree_num}). If the BZ expansion for the
location of the fixed point converges  for both observables, 
then both ${\cal O}(a_0^4)$
coefficients should be small. Thus also their difference should
also be small. And indeed, using (\ref{k0_d0_diff}) 
this difference can be calculated directly, to give:
\beq
w_3^{Bj}-w_3^{D}=-\frac{83797183}{35171328} 
+\frac{275}{107}\zeta_3 \simeq 0.70686.
\label{diff_crewther}
\eeq
Thus the Crewther relation
is consistent with a very small ${\cal O}({a_0}^4)$ coefficient in the BZ
expansions for both the D-function and the polarized Bjorken
sum rule, with the caveat that
 it is also consistent with a {\em common} large contribution
for both quantities.

Another (related) observation is that thanks to the conformal Crewther relation
 at the fixed point (\ref{crewther_conformal_2}),
the BZ coefficients of the
vacuum polarization D-function  
can be calculated directly from those of the Bjorken polarized
sum rule, and vice versa.

The BZ expansion of the D-function can be obtained 
from that of the Bjorken sum rule in two ways. In the first,
one writes down the $d_i$ coefficients in terms of the $k_i$  
using (\ref{coef_crewther}). Then one uses eq. (\ref{w_1_3}) to
calculate the BZ coefficients, where one finds that all the $S_i$
terms just cancel out, to obtain:
\begin{eqnarray}
w_1^{D}&=&w_1^{Bj}+\frac34 C_f \\ \nonumber
w_2^{D}&=&w_2^{Bj}+\frac32 w_1^{Bj} C_f+\frac{9}{16}C_f^2 
\end{eqnarray}
 The second way is by first calculating the BZ expansion for
the Bjorken sum rule, using eq. (\ref{w_1_3}), as in
(\ref{BZ_BjSR}), and then substituting the entire series
into the conformal Crewther relation (\ref{crewther_conformal_2}), and
finally expanding the rational polynomial again in terms of $a_0$. The two
methods are equivalent order by order.  

The practical implication of this is the following: if the BZ series
indeed converges fast, then the rational polynomial one gets by
substituting the BZ series for the D-function in the conformal
Crewther relation (\ref{crewther_conformal_2}) should itself be numerically 
close to the power expansion (\ref{Ree}). Its
higher-order coefficients should provide some rough estimate of the unknown
higher-order terms. We present such an analysis in fig. 7, where we
show both the straightforward BZ expansion for the vacuum polarization
function, and the result of substituting the BZ expansion for the
Bjorken sum rule in (\ref{crewther_conformal_2}).   
The caveat is that the results of this analysis could be invalidated
if there are higher-order effects that break the BZ expansion, and are  
common to the D-function and the Bjorken sum rule.

\subsection{The Derivative of the $\beta$ function}

Following ref. \cite{CaSt} we now consider the derivative
of the $\beta$ function at the fixed point, defined
by:
\beq
\gamma\equiv\left.\frac {d\beta(x)}{dx}\right\vert_{x=x_{FP}}=
-\beta_0 x_{FP}\left[2+3cx_{FP}+4c_2\left(x_{FP}\right)^2+\cdots \right]
\label{gamma_def}
\eeq

From refs. \cite{St,CaSt} we adopt the following form for the $\beta$
function:
\beq
\frac{\beta_0}{\beta(x)}=-\frac{1}{x^2}+\frac{c}{x}-
\frac{1}{\hat{\gamma}(x_{FP}-x)}+H(x)
\eeq
where $H(x)$ is a power series in the coupling $H(x)=H_0+H_1x+\cdots$,
and $\hat\gamma=\gamma/\beta_0$.
$\gamma$ is called a ``critical exponent'' since it
determines the rate at which
the coupling approaches the fixed point according to
\beq
x-x_{FP}= \left(Q^2/\Lambda^2_{\eff}\right)^{\gamma}
\eeq
where $\Lambda_{\eff}$ is the observable-dependent QCD scale \cite{CaSt}.

It is well known that $\gamma$ is independent of the renormalization
scheme, so long as the
transformations relating the different schemes are non-singular (see
ref. \cite{Chyla} and appendix B in ref. \cite{CaSt} and references therein).

It is straightforward to obtain the BZ expansion for $\hat{\gamma}$ by
substituting eq. (\ref{x_FP}) in eq. (\ref{gamma_def}) and expanding
in powers of $a_0$. The result is:
\beq
\hat\gamma=a_0\left(1+g_1a_0+g_2{a_0}^2+\cdots\right)
\eeq
where
\begin{eqnarray}
\label{g_1_3}
\lefteqn{g_1=c_{1,0}} \nonumber \\
\lefteqn{g_2=c_{1,0}^2-c_{3,-1}-c_{2,-1}^2} \nonumber \\
\lefteqn{g_3=c_{1,0}^3-2c_{4,-1}-c_{3,0}-4c_{2,-1}^3-2c_{2,0}c_{2,-1}}
\nonumber \\
&& -4c_{1,0}c_{3,-1}-5c_{1,0}c_{2,-1}^2-6c_{2,-1}c_{3,-1}
\end{eqnarray}
The coefficients $g_i$ can be proven to be universal \cite{Grunberg},
in the sense that they are the same for any physical quantity, 
in agreement with the expectation that $\gamma$ is independent of the
renormalization scheme in which the $\beta$ function is defined.

We now turn to the results. The BZ expansion for $\hat\gamma$ gives:
\begin{eqnarray}
\hat\gamma=a_0 + \frac{19}{4}{a_0}^2+\left(\frac{633325687}{105513984}
-\frac{5335}{428}\zeta_3\right){a_0}^3+ \nonumber \\
\left(\frac{43834503808535 }{
    270959910912}+\frac{590624393}{6594624}\zeta_3-2c_{4,-1}
\right){a_0}^4
\end{eqnarray}
and finally,
\beq 
\hat\gamma=a_0+4.75{a_0}^2-8.89129 {a_0}^3+
  (269.43288-2c_{4,-1}){a_0}^4
\label{gamma_num}
\eeq
According to ref. \cite{CaSt}, the 
divergence of the $\gamma$ series is not so bad to exclude the 
possibility of a perturbative
fixed point for the physical case of $a_0=0.36$ ($N_f=2$).
We doubt this assertion, since the
values the different terms in this case are: 1, 1.71, -1.15.
Even for $a_0=0.26$ ($N_f=6$) one obtains a rather slowly converging
series, with the different terms contributing as follows: 1, 1.24, -0.6.

If we use Pad\'e approximants (PA's) to estimate the unknown
${\cal O}(a_0^4)$ term, we obtain a very inconsistent result.
The $[2/1]$ PA gives:
\beq
g_3^{[1/2] PA}= -  192.494
\eeq
while the $[1/2]$ PA gives
\beq
g_3^{[2/1] PA} = 16.982.
\eeq
The difference between the two PA predictions above provides an
estimate of the uncertainty that we have assuming that $g_4$ is close
to zero. We therefore disagree with the assertion of \cite{CaSt} that
a reasonable estimate for $c_{4,-1}$ can be obtained from the
assumption that $g_3=0$.

It is interesting that the $g_i$-s are the same, not
only for physical schemes, but also for
$\MSbar$, even though the $\MSbar$ coupling is unphysical. 

\subsection{Conclusions}

The main question one would like to answer is at what $N_f$ does the BZ
expansion break down. Unfortunately, three terms in the expansion are
not quite enough to provide a definite answer. In order to measure the
convergence of the BZ expansion, we study the $N_f$ dependence of the
ratio of the ${\cal O}({a_0}^3)$ term and the partial-sum:
\beq
\frac{w_2{a_0}^3}{x_{FP}}=\frac{w_2{a_0}^3}{a_0
  +w_1{a_0}^2+w_2{a_0}^3}.
\eeq
$w_2{a_0}^3/x_{FP}$ provides some rough measure of the convergence of
a series: a divergent series where all the terms are equal and positive
yields a ratio of $1/3$ above. If the signs oscillate, it yields $1$. 
This ratio is presented in fig. 6 for the various BZ series. 
It is evident that the BZ expansion
behaves differently for different physical quantities: while the
expansion for value of the effective-charge at freezing seems to
converge for any $N_f$ for
the polarized and non-polarized Bjorken sum rules and for the vacuum
polarization D-function, 
it breaks down early for the GLS 
effective-charge (due to the light-by-light type terms),
for the hadronic Higgs decay width effective-charge, for the static
potential effective-charge  
and for the critical exponent $\gamma$. It seems that the
BZ expansion is reliable down to $N_f=12$ in all cases, 
a point we shall come back to below.

We conclude this section by comparing the picture one obtains for
perturbative freezing from the two approaches studied, namely,
finding zeros of the $\beta$ function in an `optimized scheme', and
the BZ expansion. As a representative example, we choose the
vacuum polarization D-function, and show in fig. 7 the value of the
effective charge at freezing, as calculated by the ECH and PMS methods,
together with the results from the BZ expansion. For the latter,
we show both the result of a direct calculation  (\ref{Ree_num})
and the one obtained from the BZ expansion for the Bjorken sum rule
using the conformal Crewther relation.

We interpret the difference between the ECH and PMS results, as an
intrinsic uncertainty of the `optimized scheme' approach in this
context. Similarly, we interpret the deviation
between the two BZ results as a measure of the intrinsic uncertainty
of the BZ approach, related to the fact that we are 
using a power expansion in $a_0$, rather that a more generic function
of $a_0$.

From the comparison of the two approaches for calculating $x_D^{FP}$,
i.e. ECH/PMS vs BZ, we conclude that
for $N_f\gsim 6$, the three-loop result can lead to
a perturbative fixed point -- as shown in fig. 7, 
the two methods agree,
and predict a relatively small effective coupling in the infrared
limit: $x\lsim0.3$. 
We emphasize again that a zero in the truncated ECH/PMS  $\beta$
can be easily washed out by higher order corrections.
An extreme example is provided by the Higgs decay width effective
charge, for which at the 3-loop order it seems that there is a reliable
fixed point, but in fact the perturbative series breaks down.  
 
The fact that various (space-like) effective charges run
according to a very similar RG equation (at least up to 3-loops order)
suggests that perturbative freezing will occur together and therefore
that perturbative freezing at high enough order can be indicative for
the existence of a fixed point in the full theory.  

The BZ expansions for the D-function, the polarized and the non-polarized
Bjorken sum rules show fast convergence up to order 
${\cal O}(a_0^3)$. The Crewther relation is consistent with a small
${\cal O}(a_0^4)$ coefficient for the D-function and the polarized Bjorken
sum rule, but it is also consistent with a common large contribution
at order ${\cal O}(a_0^4)$ for both quantities. From the Crewther
relation it is clear that if one of these quantities freezes, so does
the other.

An early break-down of the BZ expansion for physical quantities 
was identified for the critical exponent $\gamma$, for the static
potential, for the derivative of the Higgs decay width and for the GLS
sum rule. 

If we assume that existence of a genuine fixed point will be realized
in a perturbative manner, then we should expect convergence of the BZ
expansion for {\em any} (infra-red finite) physical quantity. Using
fig. 6, this leads to a prediction that 
\beq
N_f^{crit} \gsim 12.
\eeq 
This result agrees with the results of Appelquist 
{\em et al.} \cite{Appelquist}, 
which are based on non-perturbative calculations and also with 
lattice simulations they refer to. On the other hand, it contradicts
the results other lattice simulations \cite{Lattice}.

Although this is outside the main subject of this paper,
we emphasize again a nice feature of the perturbative
expansions in QCD, that was noticed in two different occasions in the
previous sections. 
This is the idea that the strong numerical cancelation between
different irrational numbers (in QCD these are the $\zeta_i$ terms) is usually
not accidental and most likely 
provides an indication of some yet unknown deep relation that is
encoded in the perturbative coefficients. Such a cancelation was
found in the difference between the second RG
invariants of the D-function and of the Bjorken polarized sum rule, as
well as in the ${\cal O}(a_0^3)$ term in the BZ expansion for 
the non-polarized Bjorken sum rule.

\section{Analytic continuation and time-like quantities}

\subsection{The D-function and $R_{e^+e^-}$}

In this section we concentrate on the vacuum polarization D-function
and the time-like observables that are related to it through 
dispersion relations: $R_{e^+e^-}$ and $R_\tau$.

The three loop analysis \cite{MatSt,Ree_Kat}, based on the $\beta$ 
function in an `optimized scheme', as briefly outlined
in Sections 2.1 and 2.2, 
suggests that the D-function effective charge does not freeze
for $N_f\leq4$, while the related time-like quantities do.
The differences between the values of the time-like and space-like
effective charges at freezing become significant already
at $N_f\lsim 8$.
On the other hand, in the framework of the BZ expansion, the time-like 
and space-like quantities have the {\em same} expansion for the value
of the effective charge at freezing.
In the following sections we shall examine this issue on a deeper level. We
will show that it is inconsistent to discuss perturbative freezing
of $x_R$ (or $x_\tau$) when the corresponding space-like effective
charge has a ``Landau-pole''. We will also
explain that the terms that are related to the analytical continuation 
are not supposed to change the value of the effective charge at
freezing, i.e. $x_R(0)=x_\tau(0)=x_D(0)$, {\em provided} $x_D(0)$ is well
defined. 

We start by recalling \cite{Bjorken} 
the analyticity properties of the D-function 
and the relations between the D-function and $R_{e^+e^-}$. 
From the optical theorem
\beq
R_{e^+e^-}(s)=\frac{1}{\pi}{\rm Im}\{ \Pi(-s)\}
\label{optical_theorem}
\eeq
where $\Pi(Q^2)$ is defined in (\ref{Pi_def}) and 
$s>0$ is a time-like momentum. 
From causality one expects that the only singularities of $\Pi(Q^2)$
are on time-like axis, i.e. on the negative real axis: $Q^2=-s$ with $s>0$.   
The spectral density function, $\beta_R(s)$ is defined by
\beq
\beta_R(s)=s\frac{dx_R}{ds}
\label{beta_R_def}
\eeq
Clearly $\beta_R$ is also the $\beta$ function for the coupling $x_R$.
Differentiating (\ref{optical_theorem}) one obtains a similar relation 
between $\beta_R$ and the space-like D-function effective charge 
(\ref{D_def}),
\beq
\beta_R(s)=-\frac{1}{\pi}{\rm Im}\{ x_D(-s)\}
\label{beta_R}
\eeq

Based on the above, one can express the D-function 
 as a dispersive integral over $R_{e^+e^-}$,
\beq
x_D(Q^2)=- \int_0^\infty ds\frac{\beta_R(s)}{s+Q^2}
=  Q^2\int_0^\infty ds \frac{x_R(s)}{(s+Q^2)^2}.
\label{DR}
\eeq
The relations between the coefficients of the corresponding effective
charges $x_R$ and $x_D$ (\ref{r_d_relations}) 
are directly obtainable from (\ref{DR}), as
explained in ref. \cite{Bjorken}.

The inverse relation is:
\beq
R_{e^+e^-}(s)=\frac{1}{2\pi i}\int^{s+i\epsilon}_{s-i\epsilon}ds'
\frac{d\Pi(-s')}{ds'}=-\frac{1}{2\pi
  i}\int_{s-i\epsilon}^{s+i\epsilon}ds'\frac{D(-s')}{s'},
\label{RD0} 
\eeq
where the integration contour lies in the region of analyticity of
$D(-s)$, that is, around the cut ${\rm Re}\{s'\}>0$. The contour can also be
deformed to a circle,  
\beq
R_{e^+e^-}(s) = \frac{1}{2\pi i}\oint_{\vert
  s'\vert=s}\frac{ds'}{s'}D(-s').
\label{RD1_old} 
\eeq
Relations (\ref{RD0}) and (\ref{RD1_old}) are also true for the
corresponding effective charges, i.e. replacing $R_{e^+e^-}$ and
$D(Q^2)$ by $x_R$ and $x_D$, respectively, for instance,    
\beq
x_R(s) = \frac{1}{2\pi i}\oint_{\vert
  s'\vert=s}\frac{ds'}{s'}x_D(-s').
\label{RD1} 
\eeq

It is clear from eq. (\ref{RD1}) that the infrared limit of $R_{e^+e^-}(s)$
equals that of the D-function if the latter exists:
assuming the D-function does not have an essential singularity at
$s'=0$, we have $x_D(-s')\longrightarrow x_D(0)$ and thus for small 
enough $s$ the only singularity within
the integration contour is the simple pole at the origin. From
Cauchy's theorem one then obtains $x_R(s)\longrightarrow x_D(0)$.

From this argument one learns that the infrared
limit of the exact $R_{e^+e^-}$ equals to that of the exact
D-function. But does this hold in perturbation theory? The results of
the `optimized scheme' approach and the BZ expansion indicate that
this is a delicate question. This issue is
discussed further in the following sections.

The analogous issue for $R_\tau$ will not be discussed in detail 
in this paper,
but most of our results are quite general and 
apply to it directly, since this quantity is
related to the D-function in a similar way \cite{tau}:
\beq
R_\tau(m^2_{\tau})=\frac{1}{2 \pi i} \oint_{\vert
  x\vert=1}\frac{dx}{x}\left(1-2x+2x^3-x^4\right)D(-x m_{\tau}^2)
\label{r_tau_D}
\eeq
Like in the $R_{e^+e^-}$ case, it can easily be shown that 
when $D(-x m_{\tau}^2)=D(0)$ freezes,
eq. (\ref{r_tau_D}) leads to an equality of the infrared limits
$R_\tau(0)=D(0)$.

\subsection{The analyticity structure of the D-function}

The physical quantity which can be measured directly is the time-like
$x_R$, or its derivative $\beta_R(s)$,
and not the space-like effective charge $x_D(Q^2)$.
On the other hand,  the perturbative calculation yields
$x_D(Q^2)$. Naively, there is no problem to obtain $\beta_R(s)$ from
$x_D$ via
eq.~(\ref{beta_R}). This is done by analytically continuing
the perturbative $x_D(Q^2)$ to time-like momentum
$Q^2 = {-}s$,  where $s>0$, 
and taking the imaginary part.
Alternatively, one can obtain $x_R(s)$ from eq. (\ref{RD1}).
Unfortunately, the analyticity structure of perturbative $x_D(Q^2)$ 
is inconsistent with this procedure. The 
relations between $x_D$ and $x_R$ are based on the 
assumption that $x_D(Q^2)$ is analytic in
the entire complex $Q^2$ plane, excluding the negative real axis, $Q^2<0$.
On the other hand,
a 1-loop perturbative result for $x_D(Q^2)$ violates causality, 
since it has a ``Landau-pole'' at {\em positive} real $Q^2=\Lambda_{QCD}^2\,$, \
in addition to the cut at $Q^2<0$.
In general, higher-order corrections to the $\beta$ function create
a more complicated analyticity structure which also violates causality.
Another way to see this problem is that
eqs.~(\ref{RD0}) and (\ref{RD1_old}) 
lead to different results for $x_R$ with the same $D$ function 
taken as input.
This point is discussed in ref. \cite{Analytic}.
 
In case of perturbative freezing, $x_D(Q^2)$ is
finite for any positive real $Q^2$. One might then be tempted to conclude
that freezing saves causality. This is not necessarily the case, however, since 
in principle $x_D(Q^2)$ could 
still be singular at some complex $Q^2$, while  
causality requires $x_D(Q^2)$ to be analytical in the
{\em entire} complex $Q^2$ plane, apart from the time-like axis.
Thus the resolution of the causality question requires full
knowledge of the $x_D(Q^2)$ singularity structure.
As we shall shortly demonstrate, the latter can be obtained explicitly.
  
Before investigating in general whether freezing saves causality,
we study a specific numerical
example, where we demonstrate how
perturbative freezing leads to an $x_D(Q^2)$ which is consistent
with causality. We work with a 2-loop $\beta$ function
\beq
\beta(x_D)=\frac{dx_D}{dt}=-\beta_0{x_D}^2\left(1+cx_D \right)
\label{beta_2loop}
\eeq
where $t=\ln(Q^2/\Lambda^2_{\eff})$, 
and for illustration purposes we take the hypothetical case
$\beta_0=1$ and $c=-10$. 

We will
see that the solution of (\ref{beta_2loop}) defines a unique mapping from the
entire complex $Q^2$ plane (except the time-like axis) 
into a compact domain in the complex $x_D$ plane \cite{Shifman}. 
This domain does not contain the point $x_D=\infty$. 

A straightforward integration of (\ref{beta_2loop}) yields:
\beq
\beta_0 \ln(Q^2/\Lambda^2_{\eff})=\frac{1}{x_D}
-c\ln\left(\frac{1+cx_D}{x_D}\right)
\label{2loop_int}
\eeq
In order to study the solutions of (\ref{2loop_int}) in the complex plane,
it is convenient to define: 
\beq
\ln(Q^2/\Lambda^2_{\eff}) \equiv p+i\eta, 
\label{complexQdef}
\eeq
where
$p\geq0$ and $-\pi \leq \eta < \pi$, \ and
\beq
x_D(Q^2)\equiv r(p,\eta)+ik(p,\eta) \equiv r+ik,
\label{complexX}
\eeq
where both $r$ and $k$ are real.
Eq.~(\ref{2loop_int})
can then be written as two equations for the real and imaginary parts
\beq
\beta_0 p\,=\, \frac{r}{r^2+k^2}-
\frac12c\ln\left(c^2+\frac{1+2cr}{r^2+k^2}\right)
\label{real}
\eeq
and 
\beq
\beta_0 \eta \,=\,-\frac{k}{r^2+k^2}+c
\arctan\left[k,c(r^2+k^2)+r\right]
\label{complex}
\eeq    
where $\arctan[u,v] \equiv \hbox{arg}\{u+i v\}$ takes values in $[-\pi,\pi)$.

An infrared fixed point in the two loop $\beta$ function ($c<0$) implies that
for real $Q^2 \ge 0$ eq.~(\ref{2loop_int}) has 
a real solution $0\le x_D \le ({-1}/c)$.
In the notation of eqs.~(\ref{real}) and (\ref{complex}) this means that 
for $\eta=0$ and for any $p\geq 0$ there is a solution with             
$k=0$ and $0\leq r \leq -1/c$.
Clearly, $x_D$ has a cut for 
real $Q^2 < 0$, i.e.  for $\eta=\pi$.

In order to verify that there are no other singularities in the complex 
$Q^2$ plane, we explicitly find the domain in the complex $x_D$ plane 
into which the entire $Q^2$ plane is mapped through
(\ref{2loop_int}).
This is done by taking a
contour around the cut and solving eqs.~(\ref{real}) and
(\ref{complex}) numerically for $\Lambda^2_{\eff}=1$.
We choose the following contour in the 
$Q^2$ plane:
\begin{description}
\item{a)} below the cut: $Q^2=-s-i \epsilon$, 
$0< s <\infty$, with $\epsilon=0.01$; 
\item{b)} to the right of the cut: $Q^2=p_0+i\xi$, with $p_0=10^{-10}$
 and ${-}\epsilon\le\xi\le\epsilon$;
\item{c)} above the cut:  $Q^2=-s+i \epsilon$ with $0< s <\infty$.
\end{description} 
as shown schematically in the upper part of fig.~8.
The resulting contour in the $x_D$ plane is presented in the lower part of 
fig.~8.
Clearly, the particular choice of the contour around the cut 
is arbitrary, but a
contour that is closer to the cut (smaller $\epsilon$ and $p_0$) 
would correspond to an $x_D$ domain which is only slightly larger.
We see that the entire complex $Q^2$ plane is
mapped into a  compact domain in $x_D$ plane. 
Thus there are no spurious singularities in the
complex $Q^2$ plane and so causality is preserved.

We emphasize that the solution of 
eqs.~(\ref{real}) and (\ref{complex})
described by this particular mapping is not unique and
there exist other branches of the solution.
However, none of them corresponds to a real
coupling $x_D$ along the positive real $Q^2$ axis. It is this
requirement that guarantees uniqueness of the solution.

Returning to the more general case of a 2-loop $\beta$ function, we
will now show that perturbative freezing alone ($c<0$) 
is insufficient to ensure that the
analyticity structure of $x_D(Q^2)$ is consistent with causality.
A further condition is required, namely that $\vert c\vert>\beta_0$.

The solution of the RG equation at the 2-loop order (\ref{beta_2loop})
in the complex $Q^2$ plane can be written in terms of the so-called 
Lambert $W$ function \cite{Lambert}, which is defined by 
$W(y) \exp\left[W(y)\right]=y\,$:
\beq
\begin{array}{c}
\displaystyle
x_D(Q^2)=-\frac{1}{c}\,\,\frac{1}{1+W(z)} \nonumber\\
\phantom{a}\\
\displaystyle
z = -\frac{1}{c}\exp\left(-1-\beta_0 t/c\right)
=
-\frac{1}{c\, e}
\left(\frac{Q^2}{\Lambda_{\eff}^2}\right)^{-\beta_0/c}
\end{array}
 \label{W_sol}
\eeq
$W(y)$ is a multi-valued function with an infinite number of branches,
denoted by $W_n(y)$.
We follow \cite{Lambert} as for the division of
the branches and notation. The requirement that $x_D(Q^2)$ is
real and positive for a real positive $Q^2$
(at least for $Q^2\gg \Lambda_{\eff}^2$),
is sufficient to determine the relevant branch:
for $c>0$ the physical branch is $W_{-1}(y)$, taking real values in the
range $(-\infty,-1)$, and for $c<0$ the physical branch is the
principal branch, $W_0(y)$, taking real values in the range
$(-1,\infty)$.
 
We now need to check if the singularity structure of $x_D(Q^2)$
in (\ref{W_sol}) is consistent with causality.
Let us first consider the case $c>0$:
$W_{-1}(y)$ has two branch points, at $y=0$ and at $y=-1/e$, 
which are the endpoints of two cuts stretching to $y=-\infty$. 
 
The point at $y=0$ corresponds to $Q^2\longrightarrow \infty$. The
point at $y=-1/e$ corresponds to 
\beq
Q^2_{sing}=\Lambda^2_{\eff}c^{-c/\beta_0}
=\Lambda^2_{\eff}\exp\left(-c/\beta_0\ln(c)\right)
\label{Q_sing}
\eeq
which implies that there is a ``Landau singularity'' on the space-like
axis. Thus for $c>0$ the
singularity structure is inconsistent with causality.

For $c<0$, the physical branch $W_0(y)$ has only one branch
point at $y=-1/e$, with a cut
stretching to $y=-\infty$. In the complex $Q^2$ plane this corresponds
to branch points, i.e. ``Landau singularities'',
given by (\ref{Q_sing}). 
Two different possibilities exist:
\begin{description}
\item{a) } $c<0$ and $\vert c\vert<\beta_0$. 
There is a pair of singularities in the complex $Q^2$ plane.
\item{b) } $c<0$ and $\vert c\vert>\beta_0$.  There are no
singularities in the first sheet and the perturbative solution 
is consistent with causality. In this case $x_D(Q^2)$ maps the whole complex
$Q^2$ plane into a compact domain in the coupling plane. 
The latter situation was illustrated above in the example 
with $\beta_0=1$, $c=-10$.
\end{description}
   
We will not consider here the higher-order $\beta$
function. We expect that also in this case
there are regions of parameter space resulting in
an analyticity structure consistent with causality, 
i.e. allowing the mapping of
the entire complex $Q^2$ plane into a compact domain in the complex
coupling plane.
On the basis of the experience with the two-loop $\beta$ function,
we conjecture that causality will be preserved by a wide     
subset, but not by the entire parameter space for which 
perturbative freezing holds.
In general, the perturbative result does not have the
analyticity structure implied by causality, and therefore does not
obey the dispersion relation of (\ref{beta_R}) and  (\ref{DR}). There
is a way, however, to start from the perturbative $\beta$ function and
impose the dispersion relation on it: this is the so-called Analytic
Perturbation Theory (APT) approach, which is the subject of the next section.

\subsection{Analytic perturbation theory approach}

The objective of the APT approach \cite{Analytic,dispersive} 
is to achieve the required analyticity structure, 
while retaining the correct perturbative behavior in the ultraviolet region. 

The technique is based on solving eqs. 
(\ref{real}) and (\ref{complex}) on the negative real axis, 
i.e. for $\eta=-\pi$. This can be done starting with the 
RG equation at any order. 
For instance, suppose we work at the 3-loop order and define the
coupling $x(Q^2)$ in
some arbitrary scheme and scale. Then \hbox{$x_D=x+d_1 x^2+d_2 x^3$.}
The imaginary part of $x_D(-s)$  yields the spectral density 
function $\beta_R(s)$:
\begin{eqnarray}
\beta_R(s)&=&-\frac{1}{\pi}{\rm Im}\{ x_D(-s)\}=-\frac{1}{\pi}{\rm
  Im}\left\{x(-s)+d_1x(-s)^2+d_2x(-s)^3
\right\}\nonumber \\
&=&-\frac{1}{\pi}k(t)\left[1+2d_1 r(t)
+d_2\left(3{r(t)}^2
-{k(t)}^2\right)\right] 
\end{eqnarray}  
where $x(-s)\equiv r(t)+i k(t)$, 
and $t=\ln(s/\Lambda^2_{\eff})$,\
where $s$ is time-like, $s>0$\ 
({\em cf.}~(\ref{complexX})).

The only case where the APT spectral function can be obtained in 
a closed form is for the 1-loop $\beta$ function. Substituting $c=0$ and
$\eta=-\pi$ in eqs. (\ref{real}) and (\ref{complex}), one obtains two
algebraic equations for $r(s)$ and $k(s)$, whose solutions are:
\begin{eqnarray}
\label{k_and_r}
k(t)&=&\frac{\beta_0\pi}{(\beta_0\pi)^2+(\beta_0t)^2} \nonumber \\
r(t)&=&\frac{\beta_0t}{(\beta_0\pi)^2+(\beta_0t)^2}
\end{eqnarray}
The spectral density at this order is simply 
\beq
\beta_R(t)=-\frac{1}{\pi}k(t)=
-\frac{1}{\beta_0\left[\left(\ln(s/\Lambda^2)\right)^2+\pi^2\right]}.
\label{beta_R_APT}
\eeq
Integrating the spectral density yields the time-like effective charge 
$x_R$:
\beq
x_R(t)=\frac{1}{\beta_0 \pi}\arctan\left(\pi/t \right)
+\frac{1}{\beta_0}\theta\left(-t\right).
\label{x_R_APT}
\eeq
This is a positive, continuous and monotonically increasing
function of $s$ for any $s>0$. Its infrared limit is $1/\beta_0$. 
In the ultraviolet region this function approaches the 1-loop
perturbative result $x_R(s)\longrightarrow 1/(\beta_0 t)$.

It has been recently emphasized (see the last
ref. in \cite{Analytic}) that the results of the APT approach do not
depend much on the renormalization scheme and scale. The reason for 
this stability (which has not been given in \cite{Analytic}) is
rather simple to understand already from the 1-loop result: a scale or 
scheme transformation at this order  amounts
to a shift in $\Lambda$: $\Lambda\longrightarrow \tilde\Lambda$. 
It is clear that the 
$\pi^2$ term in the denominator of (\ref{beta_R_APT})
``hides'' variations in $\Lambda$ which are not too large.

In order to study the higher order terms that make the APT result 
different from the standard perturbative 1-loop result, we construct
the expression for the spectral density $\beta_R$ (which is also the
time-like $\beta$ function) in terms of the effective-charge 
$x_R$. The perturbative regime corresponds to $s>\Lambda^2$, so
we can now ignore the $\theta$ function in the inversion of
(\ref{x_R_APT}), obtaining
\beq
t=\frac{\pi}{\tan(\beta_0\pi x_R)}
\label{tofxr}
\eeq
We then substitute (\ref{tofxr}) in (\ref{beta_R_APT}) and obtain:  
\begin{eqnarray}
\label{beta_x_R_tay}
\beta(x_R)=\frac{dx_R}{dt}&=&-\,\frac{\left(\tan(x_R\pi\beta_0)\right)^2}
{\beta_0\pi^2\left[1+\left(\tan(\beta_0\pi x_R)\right)^2\right]}\\ \nonumber
&\simeq&-\beta_0\left(x_R^2-\frac{1}{3}\pi^2\beta_0^2\, x_R^4
+\frac{2}{45}\pi^4\beta_0^4\, x_R^6 +\cdots\right) 
\end{eqnarray}
Formally we work at the 1-loop level, so the higher-order terms in 
the {\em r.h.s.} of (\ref{beta_x_R_tay}) are only a part of the 
full higher-order result. Still,
we see that solving the RG equation on the
time-like axis yields an infinite series of $\pi^2\beta_0^2$ terms.
For instance, we recognize the coefficient of the
first correction, $-\beta_0^2\pi^2/3$, as the
difference between the ECH coefficients $\rho_2^D$ and $\rho_2^R$
in eq. (\ref{rho_r_D_relations}). Thus the APT approach can be viewed
as a method of resumming the infinite series of terms associated with 
the analytical continuation.

Following \cite{Analytic} we construct the corresponding space-like
effective coupling, which is defined through the dispersion relation
(\ref{DR}). The integral can be performed analytically, and yields
\beq
x_{\APT}(Q^2)=\frac{1}{\beta_0}
\left[\frac{1}{\ln\left(Q^2/\Lambda^2\right)}
+\frac{\Lambda^2}
{\Lambda^2-Q^2}\right],
\label{removing_Landau}
\eeq
The 1-loop APT effective coupling result (\ref{removing_Landau}) 
contains a first term that is just the 1-loop perturbative result 
and a second term that exactly cancels the ``Landau-pole''. Since this 
term is a power correction, it does not alter the perturbative ultraviolet
behavior.

Let us summarize the characteristics of the APT result, which are
demonstrated above: 
by construction $x_{\APT}(Q^2)$ has a cut at ${\rm Re}\{ Q^2\}<0$ and
no other singularities in the complex plane, and in this sense it is
appropriate to describe the D-function.
Again, by construction, the APT result is consistent with the
requirement that the time-like and space-like infrared limits are
equal $x_R(0)=x_{\APT}(0)$. At 1-loop 
the infrared limit value is $x_{\APT}(0)=1/\beta_0$.

We stress that there is an important difference between perturbative
freezing, that leads to a finite infrared limit  
within perturbation theory, and the APT approach,
where $x_{\APT}(Q^2)$ is not purely perturbative, as it contains power
corrections that are due to the imposed analyticity.  
The non-perturbative nature of the APT
result is less transparent on the time-like axis,
where, as we saw, the APT approach can be viewed
as a method to resum the infinite series of terms that are related to
the analytical continuation.
All-order resummations are in general dangerous. It is well
known that the perturbative series itself has a zero radius of
convergence, and that it is even non Borel-summable (see, for
instance, \cite{Fischer}). 
Therefore, an all-order resummation of a partial series
can yield any arbitrary result. One has to be convinced that a
resummation procedure yields a result that is closer to the exact one, 
before utilizing it. It is important to understand that
analyticity alone is not enough to set the infrared limit:
one can still add further power corrections
which can alter the infrared limit without violating the 
expected analyticity structure \cite{zakharov}.

From the 1-loop APT result it is transparent that the terms 
in $\beta_R$ which are due to the analytical
continuation have nothing to do with the existence of an infrared
fixed point in QCD. 
Still, one can use the APT result to analyze the instability 
found in the predictions for 
the fixed point in the `optimized-scheme' method (Sec. 2.5).
As we saw,
the fixed point of the ``all-order" 1-loop APT result is at 
$x_R^{FP}=1/\beta_0$. On the other hand,
the function $\beta_R$ of eq. (\ref{beta_x_R_tay}), 
when truncated at order $x_R^4$, has a
zero at $x_R^{FP}=\sqrt{3}/(\pi\beta_0)$. However, when
truncated at order $x_R^6$, it does not have a non-trivial fixed
point.  
At higher-orders, 
the $\beta$ function has a non-trivial zero only when the series is 
truncated at an even power of $x_R^2$.
Moreover,
the convergence of the corresponding fixed point values to the limiting
``all-order" value
is quite slow. For instance, the deviation at order $x_R^4$ is
$45\%$, at order $x_R^8$ it is $26\%$, and at order $x_R^{12}$
it is $12\%$. 
This exercise shows that one cannot trust the value one gets by applying
the `optimized scheme' procedure to time-like quantities. 

Until now we discussed the APT approach for the 1-loop
$\beta$ function. It is interesting to see how the APT results change
when higher loops are included. 
As in Sec. 4.2, we restrict our discussion to the 2-loop case.
Starting from eq. (\ref{2loop_int}) and
performing analytical continuation by taking $Q^2=-s$ with $s>0$, we
obtain:
\beq
\beta_0 \ln(-s/\Lambda^2_{\eff})=\frac{1}{x_D(-s)}
-c\ln\left(\frac{1+cx_D(-s)}{x_D(-s)}\right).
\eeq
As mentioned in the previous section, inverting this implicit
relation is non-trivial. In fact, there is an infinite number of solutions
for $x_D(-s)$:
\beq
\begin{array}{c}
\displaystyle
x_D^{n\pm}(-s)=-\frac{1}{c}\,
\frac{1}{1+W_n(\zeta^\pm)}\\
\phantom{a}\\
\displaystyle
\zeta^\pm = -\frac{1}{c}\exp\left[-1-(t\pm i\pi)\beta_0/c\right]
\end{array}
\label{Wn_sol}
\eeq  
for any integer $n$.
As explained in detail in \cite{LamW}, uniqueness of the APT
coupling is guaranteed only if the time-like solution $x_D(-s)$ is
obtained from the physical space-like solution in such a way that
for $\vert Q^2 \vert \gg \Lambda^2$ the coupling $x_D(Q^2)$ is continuous for a
continuous change of the phase of $Q^2$.
From here one proceeds, just as in the 1-loop case, 
by taking the imaginary part 
$\beta_R=-(1/\pi){\rm Im}\{x_D(-s)\}$ and finally calculating
$x_{\APT}$ using (\ref{DR}), as in (\ref{removing_Landau}).     

While a full description of the analytic continuation from the
space-like to the time-like axis using the various branches of the 
Lambert W function is postponed to \cite{LamW}, we briefly present 
here some of the conclusions concerning the APT infrared limit.
There are two possibilities: if $c<-\beta_0<0$ (the perturbative solution
is consistent with causality) the APT coupling coincides with the
perturbative one and the infrared limit is
$x_{\APT}(0)=x_D(0)=-1/c$. Otherwise, for $c>-\beta_0$,
$x_{\APT}(0)=1/\beta_0$. Note the the latter is relevant both for
$-\beta_0<c<0$ cases, where there is a pair of complex singularities in the
complex $Q^2$ plane, and for $c>0$ cases, where there is one space-like
singularity.

\section{Conclusions}

In this work we investigated the possibility that
there is an infrared fixed point in QCD with $N_f$ flavors, 
\hbox{$0\leq N_f \leq 16\frac12$},  
that can be identified from perturbative calculation.
We examined the effective running coupling constant, defined from
several QCD observables in the `optimized scheme' approach and by 
the BZ expansion.

We showed that several different (space-like) 
effective-charges behave similarly to one another. This suggests that
freezing occurs for all of them together and therefore that perturbative
freezing may be indicative of a genuine fixed point.

In general, the ECH/PMS approach, when applied to space-like
quantities at the 3-loop order indicates a possible perturbative
fixed point for $N_f\gsim5$. It is clear, however, that the knowledge of
higher-order corrections is essential for a conclusive answer. 
While for some observables the BZ expansion 
has small coefficients suggesting that it converges down to low
$N_f$, in other physical schemes it breaks down quite early. 
Assuming that existence of a fixed point means that the BZ expansion
should converge for {\em any} physical quantity, we get a prediction
that $N_f^{crit}\gsim 12$. 

We emphasized a fundamental difference in QCD between 
the infrared behavior of quantities naturally defined for
space-like and time-like momentum.
We showed that perturbative freezing is
a necessary (but not a sufficient) condition for consistency of
perturbation theory with a causal analyticity structure.
When the infrared finite perturbative coupling has a causal
analyticity structure it coincides with the APT coupling.  

We emphasized that freezing leads to the equality of
the time-like and space-like effective couplings at the fixed point. 
In this context it is important to stress that
the terms due
to the analytical continuation in the time-like
$\beta$ function are in principle not related
to perturbative freezing. 
In the BZ expansion such terms cancel out, making it more reliable
than the
`optimized scheme' approach, where they can lead to the appearance of 
spurious infrared fixed points.

\begin{flushleft}
{\large\bf Acknowledgments}
\end{flushleft}
E.G. would like to thank G.P. Korchemsky 
for very useful discussions and to the organizers of the YKIS'97
international seminar in Kyoto, in December '97, where a part of this work
was done.
The authors also thank J.~Ellis for comments on the manuscript.
Special thanks are due to G.~Grunberg for very useful discussion
and for constructive criticism of the original version of the
paper.
This research was supported in part by the Israel
Science Foundation administered by the Israel Academy of Sciences and
Humanities, and by a Grant from the G.I.F., the German-Israeli
Foundation for Scientific Research and Development and by the Charles
Clore doctoral fellowship.

\newpage
\begin{figure}[htb]
\begin{center}
\mbox{\kern-0.5cm
\epsfig{file=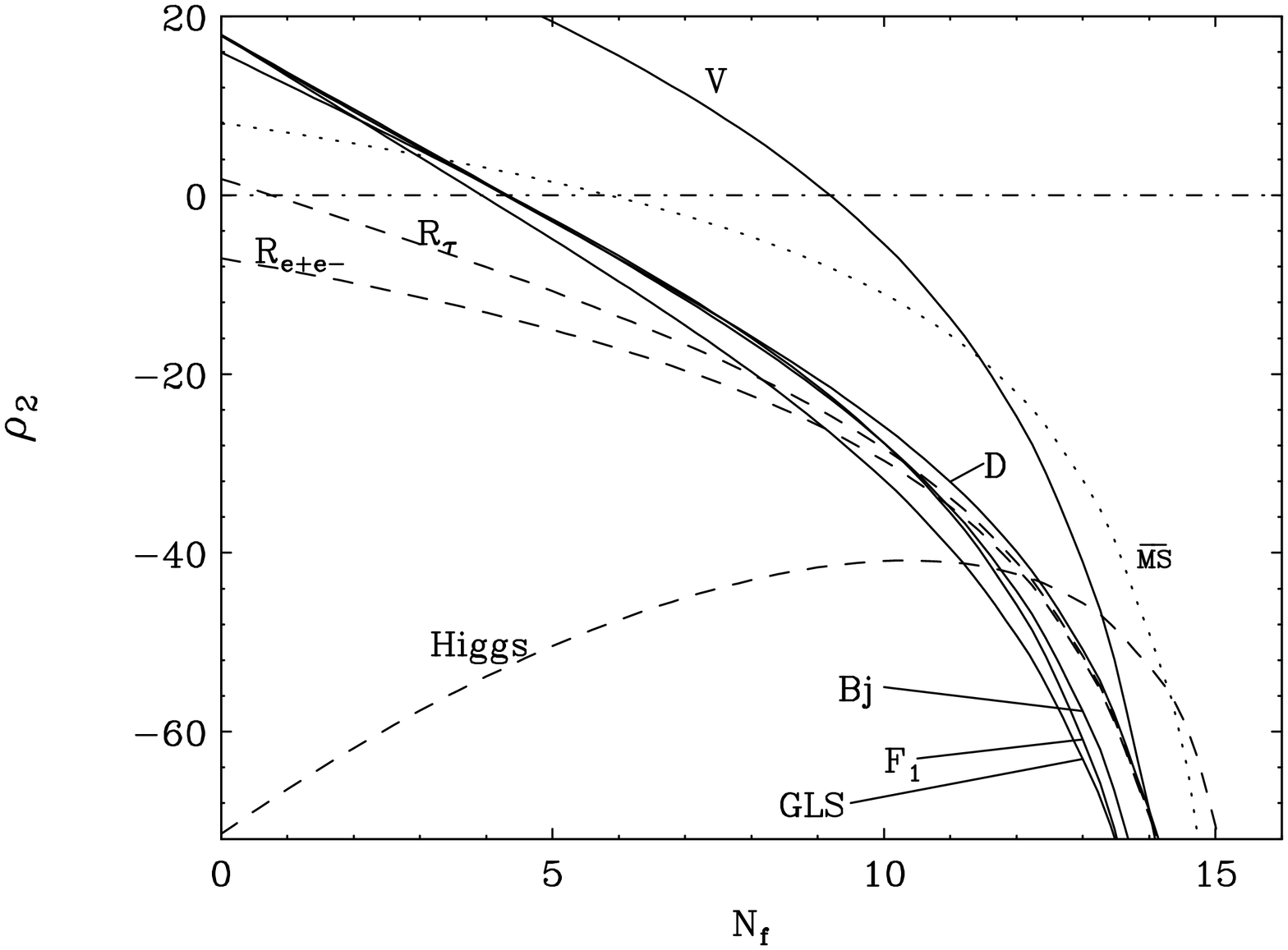,width=10.0truecm,angle=90}
}
\end{center}
\caption{The second renormalization-group invariant $\rho_2$ 
as a function of the number of light flavors for
  various space-like (continuous lines) and time-like (dashed lines) 
quantities.  
 Also shown is $c_2$, the
 second coefficient of the $\beta$ function in $\MSbar$ (dotted line). 
Note the closeness of the curves for all the space-like quantities 
except the static potential ($V$).
 }
\label{figI}
\end{figure}

\newpage
\begin{figure}[htb]
\begin{center}
\mbox{\kern-0.5cm
\epsfig{file=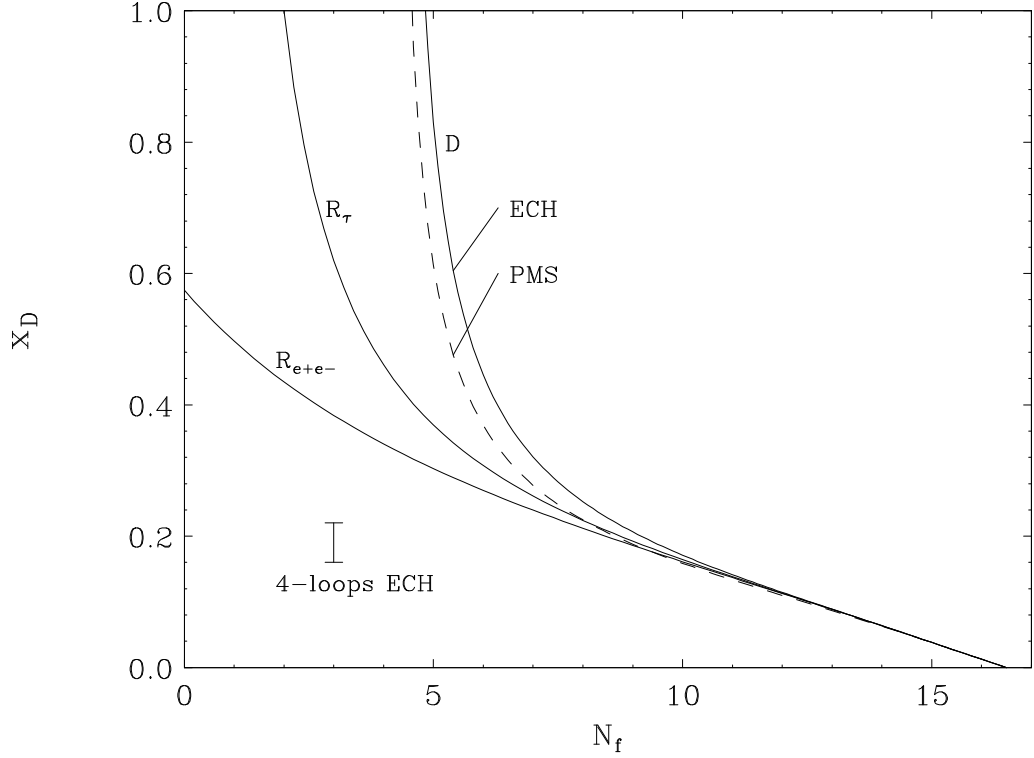,width=10.0truecm,angle=90}
}
\end{center}
\caption{The values of the effective charges at the fixed point,
as a function of $N_f$, calculated  
 in the ECH scheme (continuous lines) at
  the three loop order: upper line -- the space-like vacuum
  polarization D-function, middle line -- $R_\tau$  and lower line -- 
  $R_{e^+e^-}$. The dashed line represents the
  D-function effective charge at freezing calculated in the PMS scheme
  at the three loop order. Also presented is
the ECH result for $N_f=3$ at the four loop order as calculated using predicted
values for $d_3$ (see Sec. 2.5). 
 }
\label{figII}
\end{figure}

\newpage
\begin{figure}[htb]
\begin{center}
\mbox{\kern-0.5cm
\epsfig{file=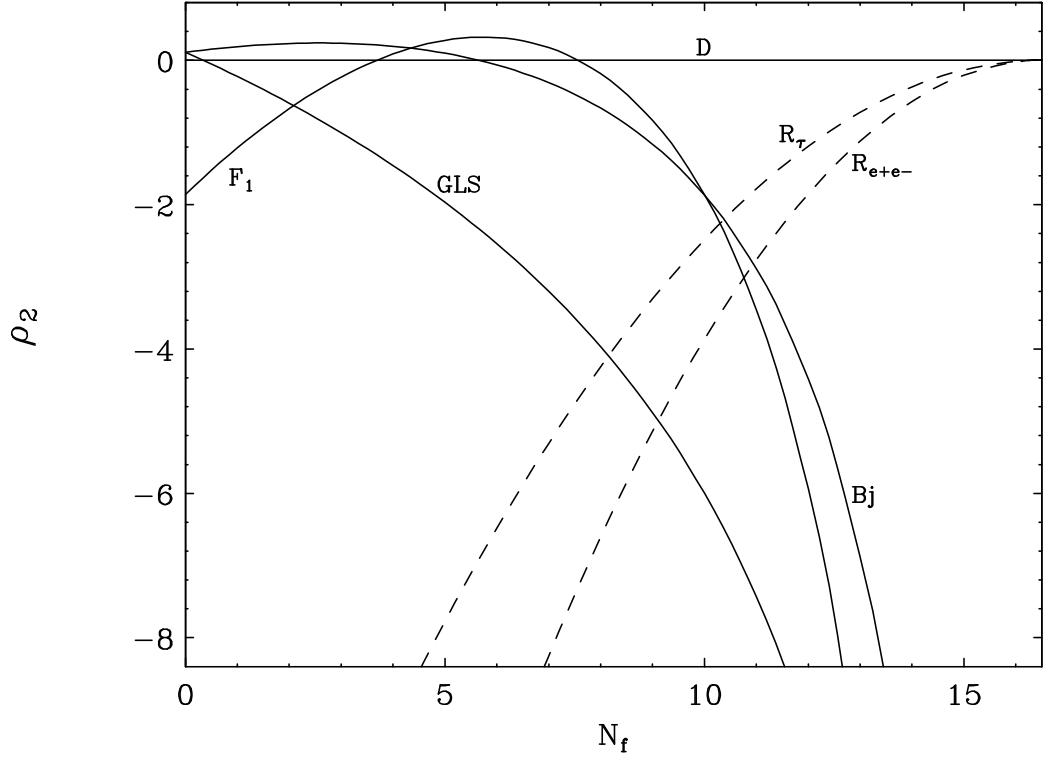,width=10.0truecm,angle=90}
}
\end{center}
\caption{
The differences between $\rho_2$ values for various
quantities and for the vacuum polarization D-function.
Space-like quantities are plotted as continuous lines, while time-like 
quantities appear as dashed lines. 
The vertical axis is magnified $10$ times compared to fig.~1.
At small $N_f$ all the space-like $\rho_2$ invariants 
are close to each other. 
For $N_f \approx 16\frac12$, 
$\rho_2^D$ coincides with those of the related time-like quantities.
 }
\label{figIII}
\end{figure}

\newpage
\begin{figure}[htb]
\begin{center}
\mbox{\kern-0.5cm
\epsfig{file=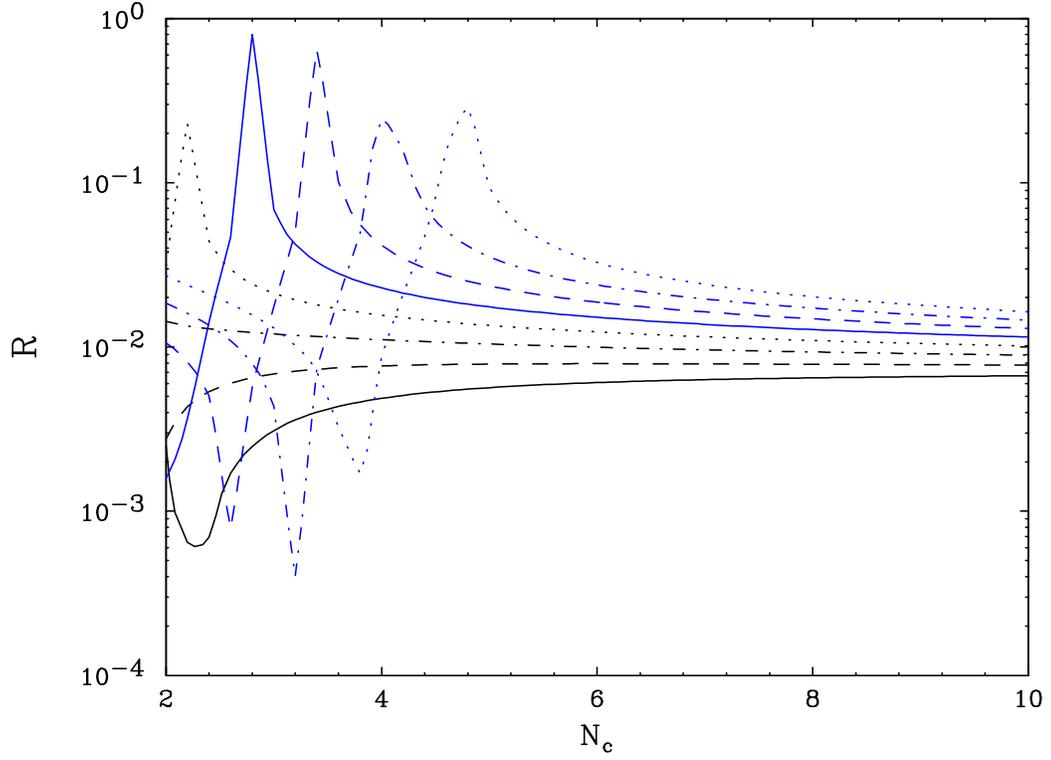,width=10.0truecm,angle=90}
}
\end{center}
\caption{The renormalized difference 
${\cal R}= \frac{\large{\vert}\rho_2^{Bj}-\rho_2^D\large{\vert}}
{\large{\vert}\rho_2^{Bj}\large{\vert}+\large{\vert}\rho_2^D\large{\vert}}$
as a function of $N_c$ for various values of $N_f$. 
The four lower lines correspond to
$N_f=0$ (continuous line), $N_f=1$ (dashed line), $N_f=2$ (dot-dash
  line), $N_f=3$ (dotted line), and the four upper curves correspond
to  $N_f=4$ (continuous line), $N_f=5$ (dashed line), $N_f=6$
(dot-dash line), $N_f=7$ (dotted line). 
 }
\label{figIV}
\end{figure}

\newpage
\begin{figure}[htb]
\begin{center}
\mbox{\kern-0.5cm
\epsfig{file=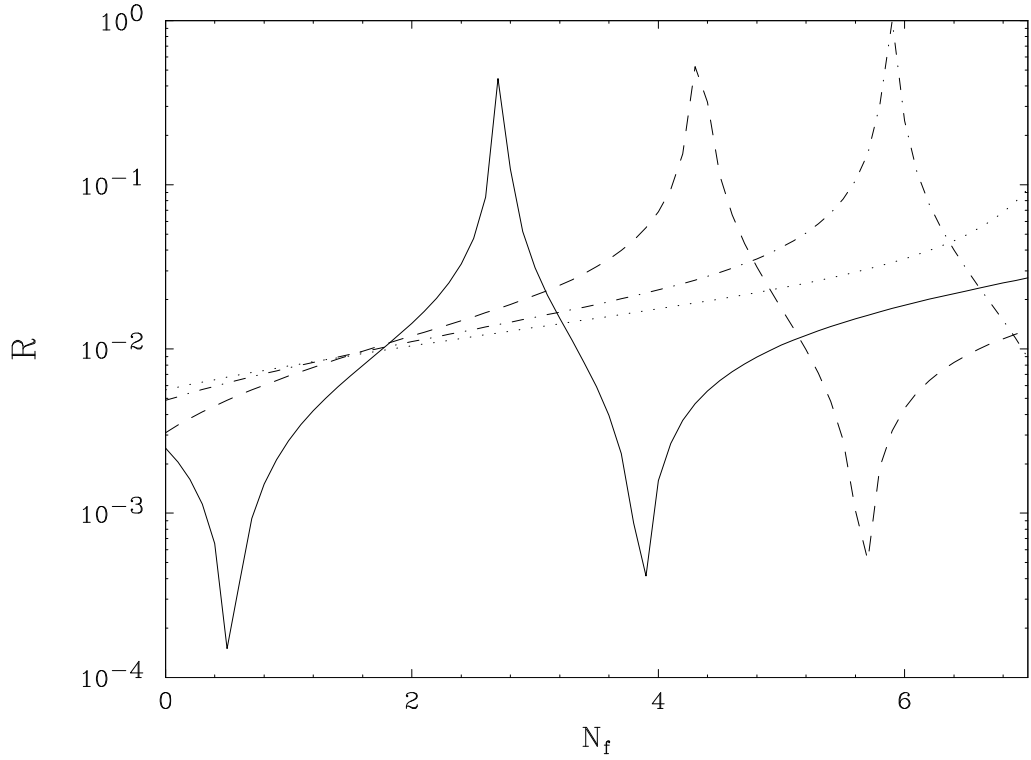,width=10.0truecm,angle=90}
}
\end{center}
\caption{
The renormalized difference 
${\cal R}= \frac{\large{\vert}\rho_2^{Bj}-\rho_2^D\large{\vert}}
{\large{\vert}\rho_2^{Bj}\large{\vert}+\large{\vert}\rho_2^D\large{\vert}}$
 as a function of $N_f$ for various values of $N_c$.
  The four lines correspond to
  $N_c=2$ (continuous line), $N_c=3$ (dashed line), $N_f=4$ (dot-dash 
line), $N_f=5$ (dotted line). 
 }
\label{figV}
\end{figure}

\newpage
\begin{figure}[htb]
\begin{center}
\mbox{\kern-0.5cm
\epsfig{file=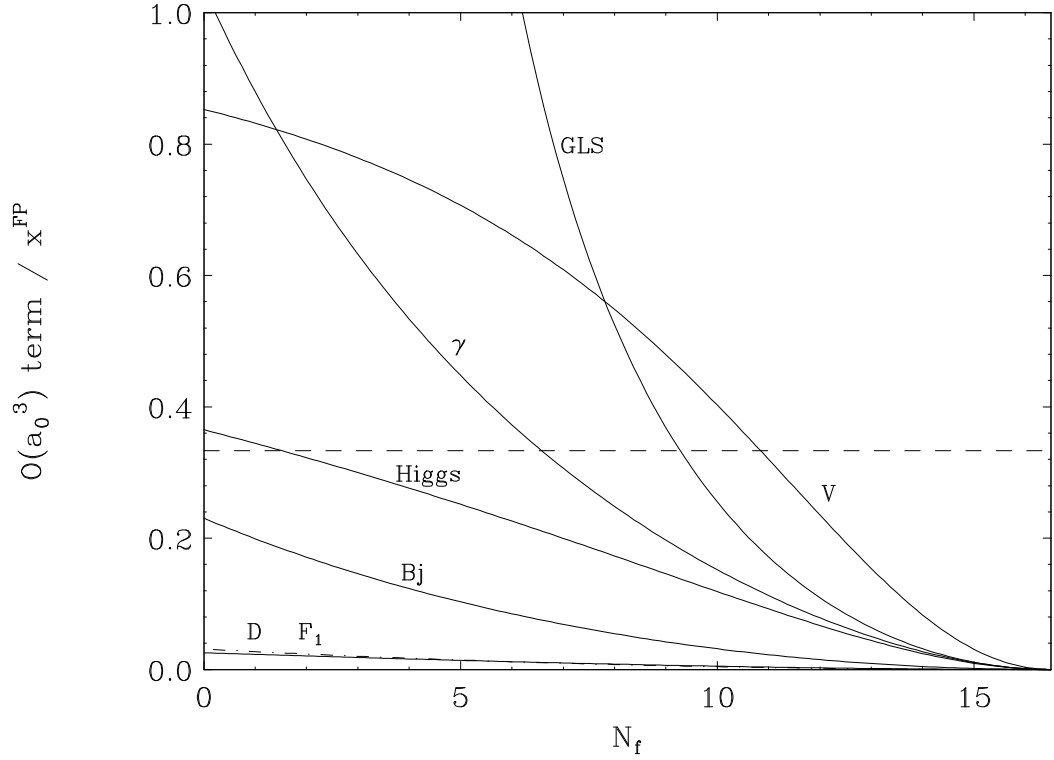,width=10.0truecm,angle=90}
}
\end{center}
 \caption{
  The ratio of the ${\cal O}({a_0}^3)$ term in the BZ expansion and the
 partial-sum ($w_2{a_0}^3/x_{FP}$) is shown for various
 quantities at the fixed point as a function of $N_f$. 
 The dot-dashed line represents the non-polarized Bjorken sum rule 
 effective charge.
 }
\label{figVI}
\end{figure}

\newpage
\begin{figure}[htb]
\begin{center}
\mbox{\kern-0.5cm
\epsfig{file=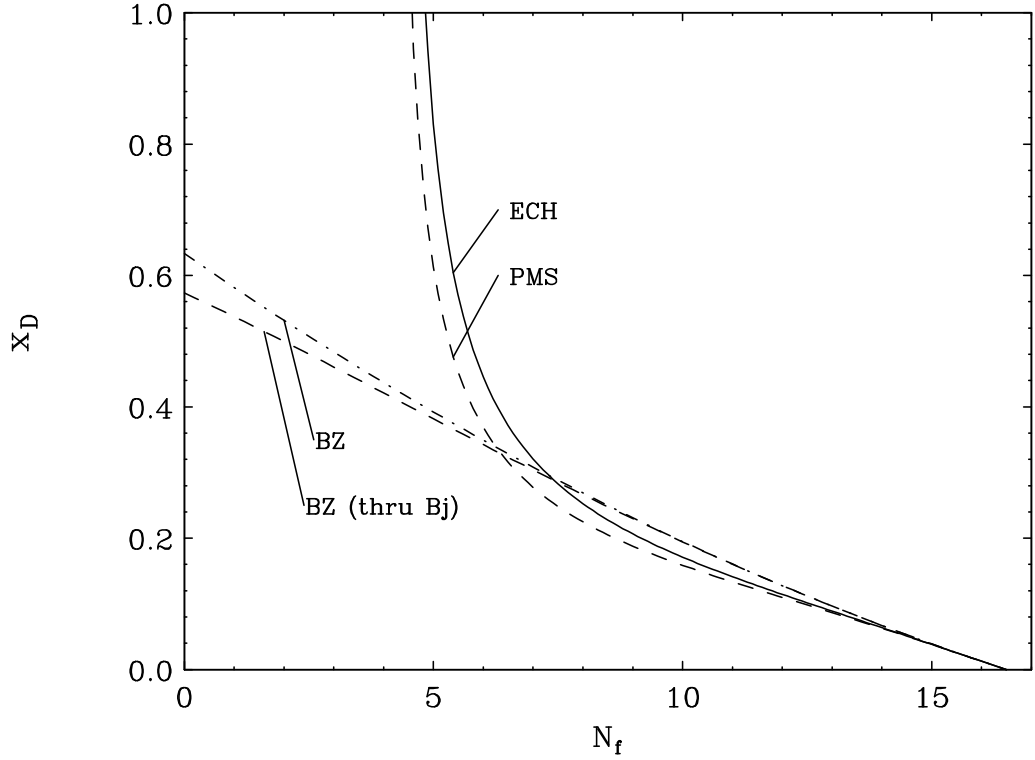,width=10.0truecm,angle=90}
}
\end{center}
\caption{ $x_{FP}^D$ as
 calculated from the ECH method (continuous line) the PMS method 
(dashed line) and the BZ expansion at
 the three loop order. For the BZ expansion, both the results of a direct
 calculation (dot-dash line) and a calculation that uses the Bjorken
 sum rule  and the Crewther relation (\ref{crewther_conformal_2}) 
(dashed line) are presented. 
}
\label{figVII}
\end{figure}

\newpage
\begin{figure}[htb]
\begin{center}
\mbox{\kern-0.5cm
\epsfig{file=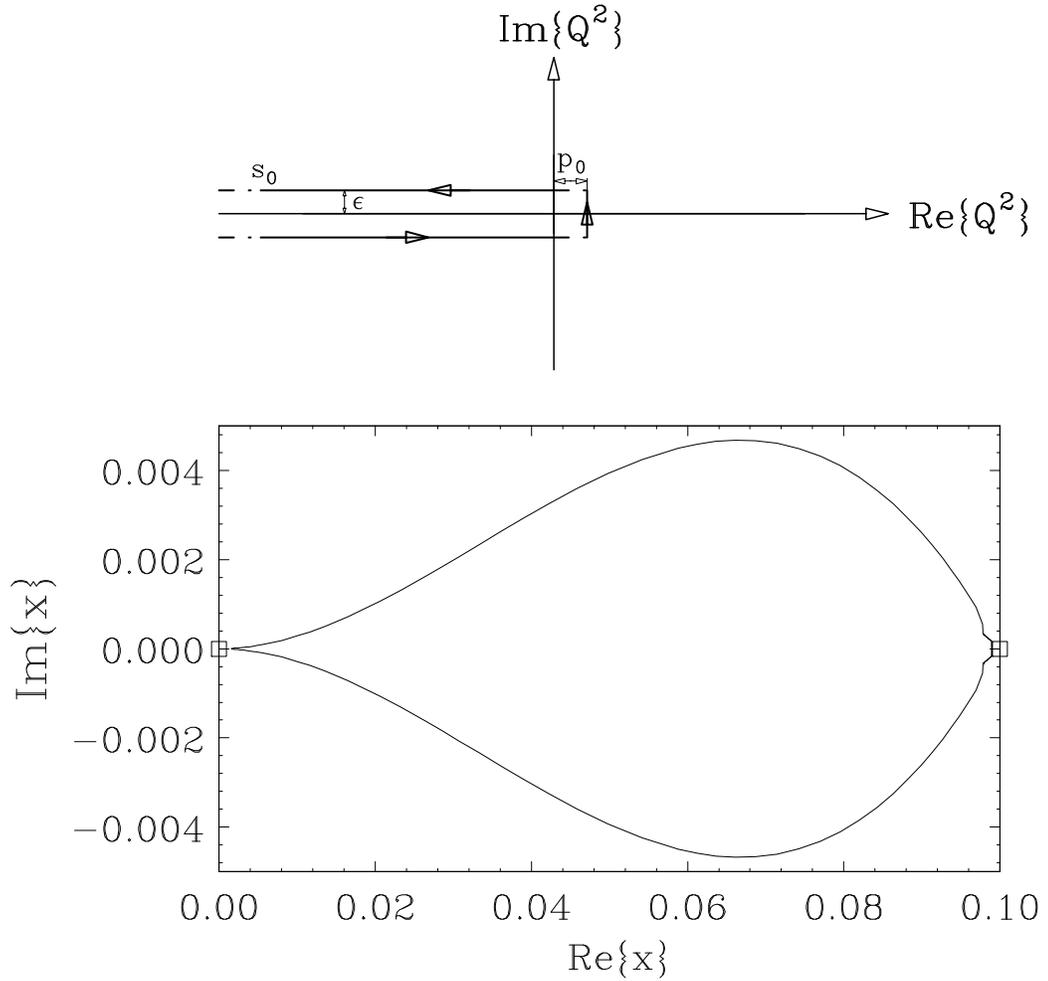,width=13.0truecm,angle=90}
}
\end{center}
\caption{ 
The mapping of the entire complex $Q^2$ plane (except the negative real axis)
into a compact domain in the complex coupling plane in the case of a
2-loop $\beta$ function (\ref{beta_2loop}),
with $\beta_0=1$ and $c=-10$.
The upper plot shows schematically the contour around the cut in the $Q^2$ 
plane as described in the text (Sec. 4.2). The lower plot shows
the image of this contour in the complex coupling ($x_D$) plane.
The two fixed points (ultraviolet, at $x_D=0$, and infrared, 
at $x_D=-1/c=0.1$) are denoted as squares.   
 }
\label{figVIII}
\end{figure}


\begin{thebibliography}{99}
\bibitem{Shifman} The authors are in debt to M. Shifman for his help on
  this matter.
\end{thebibliography}
\end{document}